\documentclass[12pt]{amsart}
\usepackage{color}

\usepackage{graphicx}
\usepackage{cite}

\newcommand{\RR}{{\mathbb R}}

\renewcommand{\Im}{\mathop{\rm Im}\nolimits}

\newtheorem{thm}{Theorem}
\newtheorem{prop}{Proposition}[section]

\newtheorem{defn}[prop]{Definition}

\def\R{\mathbb R}

\def\exp{e^}

\def\Im{\,\mathrm{Im}\,}

\def\be{\begin{eqnarray*}}
\def\ee{\end{eqnarray*}}
\def\ben{\begin{eqnarray}}
\def\een{\end{eqnarray}}

\newcommand{\sinc}{\mathrm{sinc}}
\newcommand{\diag}{\mathrm{diag}}
\newcommand{\abs}[1]{\left\vert#1\right\vert}
\newcommand{\norm}[1]{\left\Vert#1\right\Vert}
\newcommand{\inner}[2]{\left\langle #1,#2\right\rangle}
\newcommand{\paren}[1]{\left(#1\right)}
\newcommand{\bracket}[1]{\left[#1\right]}
\newcommand{\set}[1]{\left\{#1\right\}}

\numberwithin{equation}{section}
\numberwithin{prop}{section}

\def\squarebox#1{\hbox to #1{\hfill\vbox to #1{\vfill}}}

\title
[A system of ODEs for a Perturbation of a Minimal Mass Soliton]
{A system of ODEs for a Perturbation of a Minimal Mass Soliton}

\author[J. Marzuola]
{Jeremy Marzuola}
\email{marzuola@math.uni-bonn.de}

\author[S. Raynor]
{Sarah Raynor}
\email{raynorsg@math.wfu.edu}

\author[G. Simpson]
{Gideon Simpson}
\email{simpson@math.toronto.edu}

\address{Mathematics Institute, Bonn University \\
Endenicher Allee 60, D-53115 Bonn, Germany}

\address{Mathematics Department, Wake Forest University \\
P.O. Box 7388, 127 Manchester Hall, Winston-Salem, NC, 27109 USA}

\address{Mathematics Department, University of Toronto \\
40 St. George St., Toronto, Ontario, Canada M5S 2E4}

\begin{document}

%%%%%% bibliography settings %%%%%%%%%%%%%%%
%\bibliographystyle{amsalpha}
\bibliographystyle{plain}
%\bibstyle{plain}

\begin{abstract}
We study soliton solutions to a nonlinear Schr\"{o}dinger equation with a saturated nonlinearity.  Such
nonlinearities are known to possess minimal mass soliton solutions.  We consider a small perturbation of a minimal mass soliton, and identify a system of ODEs similar to those from \cite{CP}, which model the behavior of the perturbation for short times.  We then provide numerical evidence that under this system of ODEs there are two possible dynamical outcomes, which is in accord with the conclusions of \cite{pelinovsky1996nto}.  For initial data which supports a soliton structure, a generic initial perturbation oscillates around the stable family of solitons.  For initial data which is expected to disperse, the finite dimensional dynamics follow the unstable portion of the soliton curve.
\end{abstract}

\maketitle

\section{Introduction}
We consider the initial value problem for the nonlinear Schr\"{o}dinger equation (NLS) in $\R^d \times \R^+$:
\begin{equation}
\label{nls} \left\{ \begin{gathered}
i u_t + \Delta u + g (|u|^2) u =  0 , \\
u(x,0) =  u_0 (x) ,
\end{gathered} \right.
\end{equation}
where the nonlinearity $g(s)$ is a \emph{saturated} nonlinearity of the form
\begin{equation}
\label{sat:eqn1}
g (s) = s^{\frac{q}{2}} \frac{s^{\frac{p-q}{2}}}{1 + s^{\frac{p-q}{2}}},
\end{equation}
where $2+\frac{4}{d-2} > p > 2+\frac{4}{d} > \frac{4}{d}  > q > 0$ for $d \geq
3$ and $\infty > p > 2+\frac{4}{d} > \frac{4}{d} > q > 0$ for $d <
3$.  For $|u|$ large, \eqref{nls} behaves as though it were $L^2$ subcritical
while for $|u|$ small, it behaves as though it were $L^2$ supercritical.  This guarantees both existence of soliton solutions and global well-posedness in $H^1$.

For our purposes, $p$ must be chosen substantially larger than the $L^2$ critical exponent, $\frac{4}{d}$, in order to allow sufficient regularity when linearizing the equation.  For our numerical analysis, we work in one spatial dimension, with the specific nonlinearity
\begin{equation}
\label{eq:sat1}
g(s) = \frac{s^3}{1+s^2}.
\end{equation}

The equation \eqref{nls} is globally well-posed in $H^1 \cap  L^2 (|x|^2)$ with the usual norm
\begin{eqnarray*}
\| u \|_{H^1 \cap L^2 (|x|^2)}^2 = \| u \|_{H^1}^2 + \| u \|_{L^2 (|x|^2)}^2 ,
\end{eqnarray*}
where $H^1$ is the usual Sobolev space with norm
\begin{eqnarray*}
\| u \|_{H^1}^2 = \| u \|_{L^2}^2 + \| \nabla u \|_{L^2}^2
\end{eqnarray*}
and $L^2 (|x|^2)$ is the weighted Sobolev space with norm
\begin{eqnarray*}
\| u \|_{L^2 (|x|^2)}^2 = \| |x| u \|_{L^2}^2  .
\end{eqnarray*}
This is commonly referred to as the space $H^1$ with finite variance.  The global well-posedness initial data in $H^1 \cup  L^2 (|x|^2)$ follows from the standard well-posedness theory for semilinear Schr\"odinger equations. Additionally, we assume that $u_0$ is spherically symmetric, which implies $u(x,t)$ is also spherically symmetric for all $t>0$.  Proofs can be found in numerous references including \cite{Caz} and \cite{SS}.

A soliton solution of \eqref{nls} is a function $u(t,x)$ of the form
\begin{equation}
u(t,x) = e^{i \omega t} \phi_\omega(x) ,
\end{equation}
where $\omega > 0$ and $\phi_\omega (x)$ is a positive, spherically symmetric, exponentially decaying solution of the equation:
\begin{equation}
\label{eqn:sol} \Delta \phi_\omega - \omega \phi_\omega + g
(\phi_\omega^2) \phi_\omega = 0.
\end{equation}
For our particular nonlinearity, for any $\omega > 0$ there is a unique solitary wave solution $\phi_\omega(x)$ to \eqref{eqn:sol}, see \cite{BeLi} and \cite{Mc}.

\begin{figure}
\includegraphics[width=4in]{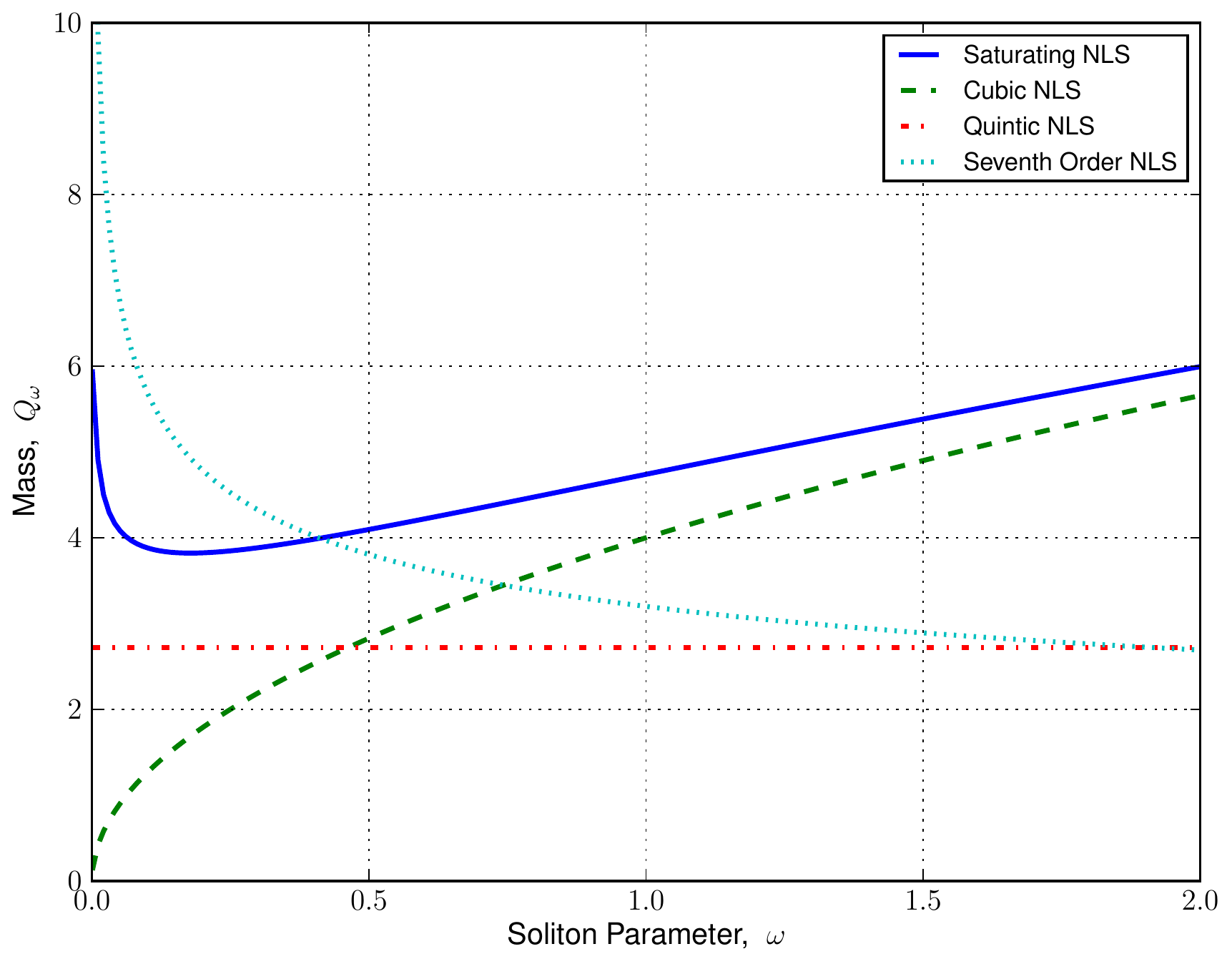}
\caption{Plots of the soliton curves ($\phi(\omega)$ with respect to
$\omega$) for a subcritical nonlinearity,
critical nonlinearity, supercritical nonlinearity, and the saturated nonlinearity \eqref{eq:sat1}.
The curves for the monomial nonlinearities are found analytically,
while the curves for the saturated nonlinearities are found
numerically using the method discussed in Section \ref{sec:sinc_disc}.  All are in $d=1$.  Here $\omega \geq .001$, as the supercritical and saturated cases diverge as $\omega \to 0$.}
\label{f:solcurves}
\end{figure}

For large $\omega$ the solitons are stable, while for small $\omega$ they are unstable. A precise stability criterion identifying stable and unstable regions is provided in \cite{GSS} and \cite{ShSt}, generalizing earlier work on stability in \cite{W1}, \cite{W2}.  This amounts to examining the relation $\omega \mapsto \|\phi_\omega\|_{L^2}^2$, defining a \emph{soliton curve}.  Where it is increasing(decreasing) as a function of $\omega$, the solitons are stable(unstable).  Several such curves appear in Figure \ref{f:solcurves}.

As can be seen numerically in Figure \ref{f:solcurves}, the nonlinearity $g$ spawns a soliton of minimal mass.  Though certain asymptotic methods can be used to describe the increasing nature of the curve as $\omega \to 0$ (multiscale methods) and $\omega \to \infty$ (variational methods), we forego an analytic description of the soliton curve and focus on the minimal mass soliton shown to exist in the numerical plot.  In \cite{CP}, Comech and Pelinovsky demonstrated that the minimal mass soliton possesses a fundamentally \emph{nonlinear} instability.  They accomplished this by finding a small perturbation that forces the solution a fixed distance away from the minimal mass soliton in finite time.  Their technique reduces to studying an ODE modeling the perturbation for short times.   For appropriate data, the ODE is unstable.

We conjecture that though the minimal mass soliton may be unstable on short time scales, on a longer time scale the solution will ultimately relax to the stable branch of the soliton curve.  This conjecture is part of a larger conjecture that solutions which do not disperse as $t \rightarrow \infty$ must eventually converge towards the stable portion of the soliton curve.  For nonlinearities with a specific two power structure, dynamics of this type were observed by Pelinovsky, Afanasjev, and Kivshar, who modelled the behavior of solutions near a minimal mass soliton by a second order ODE via adiabatic expansion in $\omega$ \cite{pelinovsky1996nto}.  By contrast, our method uses the full dynamical system of modulation parameters to find a $4$-dimensional system of ODEs which is structured to allow for eventual recoupling to the continuous spectrum.   This conjecture has also been explored numerically by Buslaev and Grikurov for two power nonlinearities in \cite{BG}, where they found that a solution which is initially a perturbation of an unstable soliton tends to approach and then oscillate around a stable soliton.

The purpose of this work is to numerically explore this conjecture.  Following \cite{CP}, we break the perturbation into the discrete and continuous parts relative to the linearization of the Schr\"{o}dinger
operator around the soliton.  The discrete portion yields a four dimensional system of nonlinear ODEs.  We further simplify the system expanding the equations in powers of the dependent variables and dropping cubic and higher terms.  

An obstacle in studying these ODEs is that the signs and magnitudes of the coefficients are not self-evident, necessitating numerical methods.  We compute these numbers, which are intimately related to the minimal mass soliton, using the $\sinc$ spectral method.  The use of the $\sinc$ function for numerically solving differential equations dates to Stenger \cite{stenger1979sgm}.  It has been successfully used in a wide variety of linear and nonlinear, time dependent and independent, differential equations, \cite{alkhaled2001sns,bellomo1995sni,bialecki1991scm,carlson1997scm
,elgamel2007san,lund1992smq,revelli2003sci,stenger2000ssn}.  In this work, we first numerically solve \eqref{eqn:sol} for the soliton as a nonlinear collocation problem.  We then use this information to compute the generalized kernel of the operator after linearization about the soliton.

With these coefficients in hand,  we numerically integrate the ODE system, plotting the results.  We find that there are two different types of behavior for the finite dimensional system, depending on the initial data.  If the initial data represents a solution which our nonlinear solver indicates can support a soliton, then we find that the solution is oscillatory.  It is initially attracted to the stable side of the curve, and, over intermediate time scales, proceeds to oscillate around the minimal mass soliton.  If we initialize with this type of data but with the unstable conditions found in \cite{CP}, the ODEs initially move in the unstable direction but quickly reverse, before commencing oscillation.  On the other hand, if we begin with initial conditions which are expected to disperse as $t \rightarrow \infty$, our data indicate that the finite dimensional dynamics push the solution along the unstable soliton curve towards the value $\omega = 0$ rather quickly.  This solution matches well to the solution for \eqref{nls} with corresponding initial data for as long as the mass conservation of the solution allows, after which our model continues to follow the unstable soliton curve but the actual solution disperses.  In \cite{pelinovsky1996nto}, the authors observed similar dynamics, with both oscillatory and dispersive regimes.

These ODEs are an approximation valid on a short time interval.  This study is the beginning of an analysis to show that perturbations of the minimal mass soliton are attracted to the stable side of the soliton curve.  In a forthcoming work we hope to show how the continuous-spectrum part of the perturbation interacts with the discrete-spectrum perturbation.  Based on the work of Soffer and Weinstein,\cite{SofWei} we expect coupling to the continuous spectrum to cause radiation damping, which will ultimately cause the solution to have damped oscillations and select a soliton on the stable side of the curve.

This paper is organized as follows.  In section 2, we introduce preliminaries and necessary definitions.  In section 3, we derive the system of ODEs.  In section 4, we explain our numerical methods for finding the coefficients of the ODEs.  In section 5, we show the numerical solutions of the ODEs and explain our results.  Finally, in section 6 we present our conclusions and plans for future work.  An appendix contains details of our numerical method for computation of the soliton and related coefficients.

{\sc Acknowledgments}
This project began out of a conversation with Catherine Sulem and JM.  JM was partially funded by an NSF Postdoc at Columbia University and a Hausdorff Center Postdoc at the University of Bonn.  In addition, JM would like to thank the University of North Carolina, Chapel Hill for graciously hosting him during part of this work.  SR would like to thank the University of Chicago for their hospitality while some of this work was completed.  GS was funded in part by NSERC.  In addition, the authors wish to thank Dmitry Pelinovksy, Mary Pugh, Catherine Sulem, and Michael Weinstein for helpful comments and suggestions.

\section{Definitions and Setup}

For data  $u_0 \in H^1 \cap L^2 (|x|^2)$,  there are several conserved quantities. Particularly important invariants are:

\begin{description}
\item[Conservation of Mass (or Charge)]
\begin{equation*}
Q (u) = \frac{1}{2} \int_{\R^d} |u|^2 dx = \frac{1}{2} \int_{\R^d}
|u_0|^2 dx.
\end{equation*}
\item[Conservation of Energy]
\begin{equation*}
E(u) = \int_{\R^d} | \nabla u |^2 dx - \int_{\R^d} G(|u|^2) dx = \int_{\R^d}
| \nabla u_0 |^2 dx - \int_{\R^d} G(|u_0|^2) dx,
\end{equation*}
where
\begin{equation*}
G(t) = \int_0^t g(s) ds.
\end{equation*}
\end{description}
%Finally, we also have the pseudoconformal conservation law:
%\begin{eqnarray}
%\| (x + 2 i t \nabla ) u \|^2_{L^2} - 4 t^2 \int_{\R^d} G(|u|^2) dx = \| x \phi \|^2_{L^2} - \int_0^t \theta (s) ds,
%\end{eqnarray}
%where
%\begin{eqnarray*}
%\theta (s) = \int_{\R^d} (4 (d+2) G(|u|^2) - 4 d g(|u|^2) |u|^2) dx.
%\end{eqnarray*}
%Note that $(x + 2 i t \nabla )$ is the Hamilton flow of the linear Schr\"odinger equation,
%so the above identity relates how the solution to the nonlinear equation is effected by the linear flow.
Detailed proofs of these conservation laws can be easily arrived at by using energy estimates or Noether's Theorem, which relates conservation laws to symmetries of an equation.  See \cite{SS} for details.

With this type of nonlinearity, it is known that soliton solutions to NLS exist and are unique.  Existence of solitary waves for nonlinearities of the type \eqref{sat:eqn1} is proved by in \cite{BeLi} in $\RR^1$ using ODE techniques and in higher dimensions by minimizing the functional
$$T(u) = \int | \nabla u |^2 dx$$
with respect to the functional
$$V(u) = \int [ G(|u|^2) - \frac{\omega}{2} |u|^2 ] dx.$$
Then, using a minimizing sequence and Schwarz symmetrization, one infers the existence of the nonnegative, spherically symmetric, decreasing soliton solution.  Once we know that minimizers are radially symmetric, uniqueness can be established via a shooting method, showing that the desired soliton occurs at only one initial value, \cite{Mc}.

Of great importance is the fact that $Q_{\omega} := Q(\phi_{\omega})$ and $E_{\omega} := E(\phi_{\omega})$ are differentiable with respect to $\omega$.  This can be determined from the works of Shatah, namely \cite{Sh1}, \cite{Sh2}.  Differentiating \eqref{eqn:sol}, $Q$ and $E$ all with respect to $\omega$, we have the relation
\begin{equation*}
\partial_{\omega} E_{\omega} = - \omega \partial_{\omega} Q_{\omega}.
\end{equation*}
Numerics show that if we plot $Q_{\omega}$ with respect to $\omega$ for the saturated nonlinearity, the soliton curve goes to $\infty$ as $\omega$ goes to $0$ or $\infty$ and has a global minimum at some $\omega = \omega_\ast > 0$; see Figure \ref{f:solcurves}.  This will be explored in detail in a subsequent numerical work by Marzuola \cite{Mnum}.

We are interested in the stability of these explicit solutions under perturbations of the initial data.
\begin{defn} The soliton is said to be {\bf{orbitally stable}} if, $\forall \epsilon >0$, $\exists \delta > 0$ such that, for any initial data $u_0$ such that $\|u_0 - \phi_\omega\|< \delta$, for any $t < 0$, there is some $\theta \in \R$ such that $\|u(x,t)-e^{i\theta}\phi_\omega(x)\| < \epsilon$.
\end{defn}
\begin{defn} The soliton is said to be {\bf{asymptotically stable}}, if, $\exists \delta >0$ such that if $\|u_0 -\phi_\omega\| < \delta$, then for large $t$, $\exists \tilde{\omega}, \tilde{\theta} > 0$ such that $u(x,t)-e^{i\tilde{\omega}t + \tilde{\theta}}\phi_{\tilde{\omega}}(x)$ disperses as a solution to the corresponding linear problem would.
\end{defn}

Variational techniques developed in \cite{W1}, \cite{W2} and  generalized to an abstract setting in \cite{GSS} and \cite{ShSt} tell us that when $\delta ( \omega ) = E_{\omega} + \omega Q_{\omega}$ is convex, or $\delta '' (\omega) > 0$, we are guaranteed stability under small perturbations, while for $\delta '' (\omega) < 0$ we are guaranteed that the soliton is unstable under small perturbations. For brief reference on this subject, see Chapter 4 of \cite{SS}.  For nonlinearities that are twice differentiable at the origin and of monomial type at infinity (which would include our saturated nonlinearities), asymptotic stability has been studied for a finite collection of strongly orbitally stable solitons by Buslaev and Perelman\cite{BP}, Cuccagna\cite{Cuc}, and Rodnianski, Soffer and Schlag\cite{RSS}.

At a minimum of $Q_{\omega}$, soliton instability is more subtle, because it is due solely to nonlinear
effects.  See \cite{CP}, where this purely nonlinear instability is proved to occur by reducing the behavior of the discrete part of the spectrum to an ODE that is unstable for certain initial conditions.

\subsection{Linearization about a Soliton}\label{lin}

Throughout this section, we use vector notation, $\vec{\cdot}$, to represent complex functions.  Any function written without vector notation is assumed to be real.   For example, the complex valued scalar function $u+iv$ will be written $\begin{pmatrix} u \\
v \end{pmatrix}$.  In this notation, multiplication by $i$ is represented by the matrix $J=\begin{pmatrix}
0 & -1 \\
1& 0
\end{pmatrix}$.  We denote by $\vec{\phi}_\omega$ the complex vector  $\begin{pmatrix}
\phi_\omega\\
0
\end{pmatrix}$, where $\phi_\omega$ is the real profile of the soliton with parameter $\omega$.
For simplicity, we suppress the $\omega$ subscript, writing $\phi$ in place of $\phi_\omega$.

For later reference, we now explicitly characterize the linearization of NLS about a soliton solution.
First consider the linear evolution of the perturbation of a soliton via the ansatz:
\begin{equation}\label{firstansatz}
\vec{u} = e^{J \omega
t}(\vec{\phi}_{\omega}(x) + \vec{\rho}(x,t))
\end{equation}
 with $\vec{\rho}= \begin{pmatrix}
\rho_{Re}\\
\rho_{Im}
\end{pmatrix}$.
For the purposes of finding the linearized hamiltonian at $\phi_{omega}$ we do not need to allow the parameters $\theta$ and $\omega$ to modulate, but when we develop our full system of equations in Section \ref{sec:odeder} parameter modulation will be taken into account.  Inserting \eqref{firstansatz} into the equation we know that since $\phi$ is a
soliton solution we have
\begin{equation}\label{firstansatzeqn}
J (\vec{\rho})_t + \Delta (\vec{\rho}) -\omega \vec{\rho}= -g( \phi^2) \vec{\rho} - 2 g'( \phi^2 )  \phi^2  \begin{pmatrix}
\rho_{Re}\\
0
\end{pmatrix} + O (|\vec{\rho}|^2).
\end{equation}
(This calculation is explained in more detailed at the start of Section 3.)   Here we have used the following calculation of the nonlinear terms of the perturbation equation:
\begin{align}
(g(|\phi+\rho|^2) (\phi + \rho)-g(|\phi|^2)\phi) & =
 ( g(\phi^2 + 2 \phi \rho_{Re} + \rho_{Re}^2 +\rho_{Im}^2) (\phi + \rho_{Re} + i \rho_{Im})-g(\phi^2)\phi) \notag \\
 = g'(\phi^2) & (\rho_{Re}^2 + 2 \phi \rho_{Re}+\rho_{Im}^2) (\phi + \rho_{Re} + i \rho_{Im}) \notag \\
&+\frac12 g''(\phi^2)(\rho_{Re}^2 + 2 \phi \rho_{Re} + \rho_{Im}^2)^2(\phi + \rho_{Re} + i \rho_{Im})
+\mbox{h.o.t.s}.\label{quad}
\end{align}
The linear terms will be absorbed into the linearized operator $J{\mathcal{H}}_\omega$,  while the quadratic terms are handled explicitly; the ${\mathcal{O}}(\rho^2)$ terms in the expansion of the equation around $\phi_\omega$ will be denoted by $N(\omega,\rho)$ in the sequel.   In this work, after expansion in powers of $\rho$, we drop all terms of order greater than two.

We are interested in linearizing this equation:
\begin{equation}
\partial_t \begin{pmatrix}
\rho_{Re}\\
\rho_{Im}
\end{pmatrix} = J\mathcal{H} \begin{pmatrix}
\rho_{Re}\\
\rho_{Im}
\end{pmatrix} +\text{h.o.t.s},
\end{equation}
where
\begin{align}
\mathcal{H} &=  \begin{pmatrix}
0 & L_{-} \\
-L_{+} & 0
\end{pmatrix}\label{hamiltonian} ,\\
L_{-} &= - \Delta + \omega - g( \phi_\omega ) ,\\
L_{+} &= - \Delta + \omega - g( \phi_\omega ) - 2 g' (\phi^2_\omega) \phi_\omega^2.
\end{align}

\begin{defn}
\label{spec:defn1}
A Hamiltonian, $\mathcal{H}$ is called admissible if the following hold: \\
1)  There are no embedded eigenvalues in the essential spectrum, \\
2)  The only real eigenvalue in $[-\omega, \omega ]$ is $0$, \\
3)  The values $\pm \omega$ are non-resonant . \\
\end{defn}

\begin{defn}
\label{spec:defn2}
Let (NLS) be taken with nonlinearity $g$.  We call $g$
admissible if there exists a minimal mass soliton, $\phi_{min}$, for (NLS)
and the Hamiltonian, $\mathcal{H}$, resulting from linearization about
$\phi_{min}$ is admissible in terms of Definition \ref{spec:defn1}.
\end{defn}

The spectral properties we need for the linearized Hamiltonian equation in order to prove stability results are precisely those from Definition \ref{spec:defn1}.  However note that it is sometimes possible to numerically solve this sort of problem even if Definition \ref{spec:defn1} does not hold; see, for example, \cite{BG}.  Notationally, we refer to $P_d$ as the projection onto the finite dimensional discrete spectral subspace $D_\omega$ of $H^1 \cap L^2 (|x|^2)$ relative to $\mathcal{H}$.  Similarly, $P_c$ represents projection onto the continuous spectral subspace for $\mathcal{H}$.

In this work, we must simply assume that $g$ is an admissible nonlinearity.  However, this assumption is  justified by the observed dynamics.  Great care must be taken in studying the spectral properties of a linearized operator; although admissibility is expected to hold generically, certain algebraic conditions on the soliton structure itself must factor into the analysis, often requiring careful numerical computations.  See \cite{DeSc} as an introduction to such methods and the difficulties therein.  To this end, in the forthcoming work \cite{MSspec}, two of the authors will look at analytic and computational methods for verifying these spectral conditions.

\subsection{The Discrete Spectral Subspace}\label{xi}

We approximate perturbations of the minimal mass soliton by projecting onto the discrete spectral subspace of the linearized operator.  We now describe, in detail, the discrete spectral subspace at the minimal mass.

Let $\omega^*$ be the value of the soliton parameter at which the
minimal mass soliton occurs.  It is proved in \cite{CP} (Lemma
3.8) that the discrete specral subspace $D_\omega$ of
${\mathcal{H}}$ at $\omega^*$ has real dimension $4$.  The functions
$\vec{e}_1=\begin{pmatrix}0 \\
\phi_\omega
\end{pmatrix}$ and $\vec{e}_2=\begin{pmatrix}
\phi'_\omega\\
0
\end{pmatrix}$ are in the generalized kernel of ${\mathcal{H}}$ at every
$\omega$.  Clearly, $\vec{e}_1$ is purely
imaginary and $\vec{e}_2$ is real. In addition to $\vec{e}_1$ and $\vec{e}_2$, at
$\omega^*$ there are two more linearly independent elements of $D_\omega$,
the purely imaginary $\vec{e}_3$ and the purely real $\vec{e}_4$.

Applying \cite{CP} (Lemma 3.9), $\vec{e}_3$ and $\vec{e}_4$ can be
extended as continuous functions of $\omega$ in such a way that
$\vec{e}_3(\omega)$ is purely imaginary, $\vec{e}_4(\omega)$ is purely real.   We write
\begin{eqnarray*}
\vec{e}_3(\omega)= \begin{pmatrix}
0\\
\alpha(\omega)
\end{pmatrix}
\end{eqnarray*}
and
\begin{eqnarray*}
\vec{e}_4(\omega)= \begin{pmatrix}
\beta (\omega)\\
0
\end{pmatrix} ,
\end{eqnarray*}
with  $\alpha$ and $\beta$ real-valued functions.  The  linearized operator, restricted to this subspace, is
\begin{equation}
J\mathcal{H}(\omega)|_{D_\omega}= \left(
\begin{array}{cccc}
0 & 1 & 0 & 0\\
0 & 0 & 1 & 0 \\
0 & 0 & 0 & 1 \\
0 & 0 & a(\omega)& 0
\end{array} \right)\label{jhomega},
\end{equation}
where $a(\omega)$ is a differentiable function that is equal to $0$ at $\omega^*$.

Before proceeding to the derivation of the ODEs, it is helpful to make a minor change of basis.  Our goal is that, in the new basis, $\left\{\vec{\tilde{e}}_1, \vec{\tilde{e}}_2, \vec{\tilde{e}}_3, \vec{\tilde{e}}_4\right\}$, $\inner{\vec{\tilde{e}}_1}{\vec{\tilde{e}}_3}=0$, which will make it easier for us to compute the dual basis.  Replace $\vec{e}_3$ by
\begin{eqnarray*}
\vec{\tilde{e}}_3 & = & \vec{e}_3 - \frac{\langle \vec{e}_1, \vec{e}_3 \rangle}{\|\vec{e}_1\|^2} \vec{e}_1 \\
& = & \begin{bmatrix}
0 \\
\tilde{\alpha}
\end{bmatrix}.
\end{eqnarray*}
To preserve the relationship $\vec{e}_3 = JH_\omega \vec{e}_4$, we
need to replace $\vec{e}_4$ by
\begin{eqnarray*}
\vec{\tilde{e}}_4 & = & \vec{e}_4- \frac{\langle \vec{e}_1, \vec{e}_3 \rangle}{\|\vec{e}_1\|^2} \vec{e}_2 \\
& = & \begin{bmatrix}
\tilde{\beta} \\
0
\end{bmatrix}.
\end{eqnarray*}
To preserve the relationship $JH_\omega
\vec{e}_3=\vec{e}_2+a(\omega)\vec{e}_4$, we replace $\vec{e}_2$ by
\begin{eqnarray*}
\vec{\tilde{e}}_2 & = & \left ( 1
+ \frac{\langle \vec{e}_1, \vec{e}_3 \rangle}{\|\vec{e}_1\|^2} \right ) \vec{e}_2 \\
& = & \begin{bmatrix}
\tilde{e_2} \\
0
\end{bmatrix} .
\end{eqnarray*}
To
preserve $JH_\omega \vec{e}_2 = \vec{e}_1$, we get that $\vec{e}_1$ must be replaced
by
\begin{eqnarray*}
\vec{\tilde{e}}_1 & = & \left ( 1 + \frac{\langle \vec{e}_1, \vec{e}_3
\rangle}{\|\vec{e}_1\|^2} \right ) \vec{e}_1 \\
& = & \begin{bmatrix}
0 \\
\tilde{e_1}
\end{bmatrix}.
\end{eqnarray*}
With these substitutions, the
$JH_\omega$ matrix on $D_\omega$ remains the same and we obtain
the relationship $\inner{\vec{\tilde{e}}_3}{\vec{\tilde{e}}_1} = 0$.
From here on we will assume that we are working with this modified
basis and simply take $\vec{e}_j := \vec{\tilde{e}}_j$ for $j = 1,2,3,4$.

We will define $\vec{\xi_i}$ to be the dual basis to the revised $\vec{e_i}$ within
$D_\omega$. That is, the $\vec{\xi}_i$ are defined by $\vec{\xi}_i \in
D_\omega$ and
$$
\langle \vec{\xi}_i, \vec{e}_j \rangle =\delta_{ij}.
$$
If we make the change of basis described above, then we can
compute the $\vec{\xi}_j$ as follows. Define $D =
\|\vec{e}_2\|^2\|\vec{e}_4\|^2-\langle \vec{e}_2, \vec{e}_4 \rangle ^2$. Then:
\begin{align*}
\vec{\xi}_1 & = \frac{1}{\|\vec{e}_1\|^2}  \vec{e}_1, \notag \\
\vec{\xi}_2 & = \frac{\|\vec{e}_4\|^2}{D}\vec{e}_2 - \frac{\langle \vec{e}_2, \vec{e}_4 \rangle}{D} \vec{e}_4 ,\notag \\
\vec{\xi}_3 & = \frac{1}{\|\vec{e}_3\|^2} \vec{e}_3, \label{xi} \\
\vec{\xi}_4 & = - \frac{\langle \vec{e}_2, \vec{e}_4 \rangle}{D} \vec{e}_2 +
\frac{\|\vec{e}_2 \|^2}{D}\vec{e}_4  . \notag
\end{align*}

As with the $\vec{e}_j$'s,
\begin{equation*}
\vec{\xi}_j = \begin{bmatrix}
0 \\
\xi_j
\end{bmatrix}
\end{equation*}
for $j = 1,3$ and
\begin{equation*}
\vec{\xi}_j = \begin{bmatrix}
\xi_j \\
0
\end{bmatrix}
\end{equation*}
for $j = 2,4$ to distinguish between vectors and their scalar components.

\section{Derivation of the ODEs}
\label{sec:odeder}

To derive the ODEs we start with a small spherically symmetric perturbation of the minimal mass soliton, then project onto the discrete spectral subspace.  Here, we closely follow  \cite{CP}.

We begin with the following ansatz, which allows $theta$ and $\omega$ to modulate:
\begin{equation}
\vec{u}(t)=e^{(\int_0^t\omega(t')dt' +\theta(t))J} (\vec{\phi}_{\omega(t)} + \vec{\rho}(t)).\label{ansatz}
\end{equation}
Recall, we have assumed $u$ to be spherically symmetric, so no other modulation parameters occur.  Unlike in \cite{CP} we do not assume that the rotation variable $\theta(t)$ is identically zero, so we need to include $\theta$ modulation in our full ansatz.   Note that the derivation which follows applies for all nonlinearities in any dimension; specialization is required only to get explicit numerical results.  This model includes the full dynamical system for spherically symmetric data and is designed in such a way that coupling to the continuous spectral subspace could easily be reintroduced.  The authors plan to analyze the effect of that coupling, which is expected to be dissipative, in a future work.  

Differentiating \eqref{ansatz} with respect to $t$, we get
$$
\vec{u}_t=[(\omega + \dot{\theta})J(\vec{\phi}+\vec{\rho}) + \dot{\omega}\vec{\phi} + \dot{\vec{\rho}}]e^{i(\int_0^t\omega(t')dt' +\theta(t))} ,
$$
where we represent differentiation with respect to $t$ by $\dot{\cdot}$ and differentiation with respect to the soliton parameter $\omega$ by $\cdot'$.  Plugging the above ansatz into the equation and cancelling the phase term yields
\begin{equation}
-(\omega+\dot{\theta})(\vec{\phi}+\rho) +  \dot{\omega} J\vec{\phi'}+ J\dot{\vec{\rho}} +\Delta \vec{\phi} +\Delta \vec{\rho} + g(|\phi+\rho|^2)(\vec{\phi} + \vec{\rho})=0 .
\end{equation}
Recall that since $\phi$ is a soliton solution, $-\omega \phi + \Delta \phi + g(|\phi|^2)\phi=0$, yielding
\begin{equation}\label{ansatzequation}
-\dot{\theta}\vec{\phi}-(\omega+\dot{\theta})(\vec{\rho}) +  \dot{\omega}J \vec{\phi'}+ J \dot{\vec{\rho}} +\Delta \vec{\rho} + g(|\vec{\phi}+\vec{\rho}|^2)(\vec{\phi} + \vec{\rho})-g(\phi^2)\phi=0.
\end{equation}

We multiply by $J$, solve for $\vec{\rho}$, and simplify.  At the same time, we collect the
$\Delta \vec{\rho}$ and $-\omega \vec{\rho}$ terms with the linear portion of
$ g(|\vec{\phi}+\vec{\rho}|^2)(\vec{\phi} + \vec{\rho})-g(\phi^2)\vec{\phi}$, which yields $JH_\omega$ as defined in \eqref{hamiltonian}.  The remaining terms of the nonlinearity are at least quadratic in
$\rho$; recall that the quadratic terms are described in \eqref{quad} and denoted $N(\omega,\rho)$.

Defining $\rho_j (t)$ as the coefficient of $\vec{e}_j (t)$ in $\rho$, we have
\begin{equation*}
\vec{\rho} = \left[ \begin{array}{c}
\rho_{Re} \\
\rho_{Im}
\end{array} \right] = \rho_1 \vec{e}_1 + \rho_2 \vec{e}_2 + \rho_3
\vec{e}_3 + \rho_4 \vec{e}_4 + \vec{\rho}_c.
\end{equation*}
Then, the above calculations give us
\begin{equation}
\dot{\vec{\rho}}= JH_\omega\vec{\rho} -\dot{\theta}\begin{pmatrix}0\\ \phi\end{pmatrix} - \dot{\theta}J \vec{\rho} -\dot{\omega} \begin{pmatrix} \phi' \\ 0 \end{pmatrix} + \vec{N} (\omega,\vec{\rho}) .
\label{rhodot}
\end{equation}

Taking the inner product of \eqref{rhodot} with each of the
$\vec{\xi}_i$ as defined in Section \ref{xi}, and applying \eqref{jhomega} yields the following system:
\begin{align}
\langle \vec{\xi}_1,\dot{\vec{\rho}} \rangle & = \rho_2 -\dot{\theta} -\dot{\theta} \langle \vec{\xi}_1, J\vec{\rho} \rangle + \langle \vec{\xi}_1, \vec{N} \rangle, \notag \\
\langle \vec{\xi}_2,\dot{\vec{\rho}} \rangle & = \rho_3 -\dot{\omega} -\dot{\theta} \langle \vec{\xi}_2, J\vec{\rho} \rangle + \langle \vec{\xi}_2, \vec{N} \rangle, \notag \\
\langle \vec{\xi}_3,\dot{\vec{\rho}} \rangle & = \rho_4 -\dot{\theta} \langle \vec{\xi}_3, J\vec{\rho} \rangle + \langle \vec{\xi}_3, \vec{N} \rangle,\label{firstsystem}\\
\langle \vec{\xi}_4,\dot{\vec{\rho}} \rangle & =a(\omega) \rho_3 -\dot{\theta}
\langle \vec{\xi}_4, J\vec{\rho} \rangle + \langle \vec{\xi}_4, \vec{N} \rangle . \notag
\end{align}
From this point forward in our approximation we drop the $\vec{\rho}_c$ component as a higher order error term.  Using the product rule, we solve the left hand side for $\dot{\rho_i}$ and put the extra terms from the derivative of the operator that projects onto the discrete spectral subspace onto the right hand side.

We have, as in \cite{CP}, that \[
P_d\dot{\vec{\rho}}=\sum \vec{e}_j \dot{\rho_j} +
\dot{\omega}\sum \vec{e_i} \Gamma_{ij}\rho_j -\dot{\omega}P_dP_d'\vec{\rho}, \]
where we have implicity defined
\begin{eqnarray*}
\Gamma_{ij} = \langle \xi_i , e_j' \rangle
\end{eqnarray*}
and used that
\begin{eqnarray*}
P_d \frac{d}{dt} P_c \vec{\rho} = -
\dot{\omega} P_d P_d' \vec{\rho}.
\end{eqnarray*}

This gives
\begin{align}
\dot{\rho_1}+\dot{\theta}& =\rho_2 - \dot{\theta}\langle \vec{\xi}_1, J \vec{\rho} \rangle +\langle \vec{\xi}_1, \vec{N} \rangle +\dot{\omega} (\langle \vec{\xi}_1, P_d'\vec{\rho} \rangle - \sum \Gamma_{1j}\rho_j ) , \notag  \\
\dot{\rho_2} + \dot{\omega} & = \rho_3 - \dot{\theta}  \langle \vec{\xi}_2, J \vec{\rho} \rangle+\langle \vec{\xi}_2, \vec{N} \rangle +\dot{\omega} (\langle \vec{\xi}_2, P_d' \vec{\rho} \rangle - \sum \Gamma_{2j}\rho_j ) , \notag \\
\dot{\rho_3} & = \rho_4 - \dot{\theta} \langle \vec{\xi}_3, J\vec{\rho} \rangle+\langle \vec{\xi}_3, \vec{N} \rangle +\dot{\omega} (\langle \vec{\xi}_3, P_d' \vec{\rho} \rangle - \sum \Gamma_{3j}\rho_j ) ,  \label{secondsystem}\\
\dot{\rho_4}& =a(\omega) \rho_3 - \dot{\theta} \langle \vec{\xi}_4, J \vec{\rho} \rangle+\langle
\vec{\xi}_4 , \vec{N} \rangle + \dot{\omega} (\langle \vec{\xi}_4, P_d' \vec{\rho} \rangle -
\sum \Gamma_{4j}\rho_j ) . \notag
\end{align}
There is also coupling to the continuous spectrum through
terms such as $\langle \vec{\xi}_1, J \vec{\rho} \rangle$ which we omit.  This can be
included in the error term and is not analyzed in our finite dimensional system.

To make the system well-determined, we must introduce two orthogonality conditions.
The first is $\langle \rho, e_2 \rangle=0$, and the second is $\langle \rho, e_1 \rangle=0$.  These represent the choice of $\omega(t)$ and $\theta(t)$ respectively that minimize the size of $\rho$.  These yields $\rho_2=\dot{\rho_2}=0$, and $\rho_1=\dot{\rho_1}=0$, respectively.

The reduced system is then:
\begin{align}
\dot{\theta} & =- \dot{\theta} \langle \vec{\xi}_1, J\vec{\rho} \rangle +\langle \vec{\xi}_1,\vec{N} \rangle +\dot{\omega} (\langle \vec{\xi}_1, P_d'\vec{\rho} \rangle - \sum \Gamma_{1j}\rho_j ) , \notag  \\
\dot{\omega} & = \rho_3 - \dot{\theta}  \langle \vec{\xi}_2, J\vec{\rho} \rangle+\langle \vec{\xi}_2,\vec{N} \rangle +\dot{\omega} (\langle \vec{\xi}_2, P_d'\vec{\rho} \rangle - \sum \Gamma_{2j}\rho_j )  , \notag \\
\dot{\rho_3} & = \rho_4 - \dot{\theta} \langle \vec{\xi}_3, J\vec{\rho} \rangle+\langle \vec{\xi}_3,\vec{N} \rangle +\dot{\omega} (\langle \vec{\xi}_3, P_d'\vec{\rho} \rangle - \sum \Gamma_{3j}\rho_j ) , \label{fullsystem1} \\
\dot{\rho_4} & =a(\omega) \rho_3 - \dot{\theta} \langle \vec{\xi}_4, J\vec{\rho} \rangle+\langle
\vec{\xi}_4,\vec{N} \rangle +\dot{\omega} (\langle \vec{\xi}_4, P_d'\vec{\rho} \rangle -
\sum \Gamma_{4j}\rho_j ) .\notag
\end{align}

In \cite{CP}, the authors further reduce this system to prove there is an initial nonlinear instability.  (Note that they have a slightly different system because
they have assumed that $\theta \equiv 0$.)  We are interested in the dynamics on an intermediate time scale; thus, we retain quadratically nonlinear terms in our equations.

Our notation is as follows.  First, we have
\begin{align*}
\langle \vec{\xi}_1, J\vec{\rho} \rangle & = \langle {\xi}_1, \rho_2 \phi'+\rho_4\beta \rangle \\
& =  \rho_4 \langle \xi_1, \beta \rangle,
\end{align*}
since $\rho_2=0$.  Denote
\begin{equation*}
c_{14} = \langle {\xi}_1(\omega^*), \beta(\omega^*) \rangle,
\end{equation*}
which is the highest order term and the only one that will figure into our quadratic expansion.  Notice that this is a real inner product of functions that normally do not appear in the same component of the complex vectors, because of the $J$ in the equation.

Similarly, we have
\begin{align*}
\langle \vec{\xi}_2, J\vec{\rho} \rangle & =   \langle \xi_2, -\rho_1 \phi-\rho_3\alpha \rangle \\
& = - \rho_3\langle \xi_2, \alpha \rangle,
\end{align*}
since $\rho_1=0$.  Denote
\begin{equation*}
c_{23} = \langle \xi_2 (\omega^*), \alpha(\omega^*) \rangle,
\end{equation*}
which is again the highest order term.

Then we have
\begin{align*}
\langle \vec{\xi}_3, J\vec{\rho} \rangle & =  \langle \xi_3, \rho_2 \phi'+\rho_4\beta \rangle \\
& =   \rho_4\langle \xi_3, \beta \rangle,
\end{align*}
since $\rho_2=0$.  Denote
\begin{equation*}
c_{34} = \langle \xi_3 (\omega^*), \beta(\omega^*) \rangle.
\end{equation*}

Finally, we have
\begin{align*}
\langle \vec{\xi}_4, J\vec{\rho} \rangle & =  \langle \xi_4, -\rho_1 \phi-\rho_3\alpha \rangle \\
& = -\rho_3\langle \xi_4, \alpha \rangle,
\end{align*}
since $\rho_1=0$.  Denote by
\begin{equation*}
c_{43} = \langle {\xi}_4(\omega^*), \alpha(\omega^*) \rangle.
\end{equation*}

We also write $g_{ij}$ for the term $\Gamma_{ij}(\omega)=\langle \vec{\xi}_i, \vec{e}_j' \rangle$ at $\omega = \omega^\ast$.

Next, consider the terms $\langle \vec{\xi}_j, P_d'\vec{\rho} \rangle $.  These terms are the $e_j$ components of $P_d'\vec{\rho}$.  We have:
\begin{align*}
P_d P_d' \vec{\rho} & = P_d \left[ \sum_{j=1}^4 \langle \vec{\xi}_j', \vec{\rho} \rangle \vec{e}_j  + \sum_{j=1}^4 \langle \vec{\xi}_j, \vec{\rho} \rangle \vec{e}_j' \right]  \\
& =\sum_{j=1}^4 \sum_{k=3}^4 \langle \vec{\xi}_j', \vec{e}_k \rangle\rho_k \vec{e}_j + \rho_3 P_d \vec{e}_3' + \rho_4 P_d \vec{e}_4'\\
& = \sum_{j=1}^4 \sum_{k=3}^4 \langle \vec{\xi}_j', \vec{e}_k \rangle\rho_k \vec{e}_j + \rho_3 (\Gamma_{13} \vec{e}_1 + \Gamma_{33} \vec{e}_3) + \rho_4 (\Gamma_{24} \vec{e}_2 + \Gamma_{44} \vec{e}_4)   \\
& =   \langle \vec{\xi}_1', \vec{e}_3 \rangle \rho_3 \vec{e}_1 + \langle \vec{\xi}_2', \vec{e}_4 \rangle \rho_4 \vec{e}_2 + \langle \vec{\xi}_3', \vec{e}_3 \rangle \rho_3 \vec{e}_3 + \langle \vec{\xi}_4', \vec{e}_4 \rangle \rho_4 \vec{e}_4 \\
&   \rho_3 (\Gamma_{13} \vec{e}_1 + \Gamma_{33} \vec{e}_3) + \rho_4 (\Gamma_{24} \vec{e}_2 + \Gamma_{44} \vec{e}_4)\\
& =  (\langle \vec{\xi}_1', \vec{e}_3 \rangle + \Gamma_{13} ) \rho_3 \vec{e}_1 + (\langle \vec{\xi}_2', \vec{e}_4 \rangle + \Gamma_{24}) \rho_4 \vec{e}_2 \\
& +(\langle \vec{\xi}_3', \vec{e}_3 \rangle + \Gamma_{33}) \rho_3 \vec{e}_3 + (\langle \vec{\xi}_4', \vec{e}_4 \rangle + \Gamma_{44}) \rho_4 \vec{e}_4 .
\end{align*}

%\begin{align*}
%P_d \frac{d}{dt} (P_c \vec{\rho}) & =  P_d \frac{d}{dt} \left[ \vec{\rho} - \sum_{j=1}^4 \langle \vec{\xi}_j, \vec{\rho} \rangle \vec{e}_j \right]  \\
%& =  P_d (P_c \frac{d}{dt} \vec{\rho}) - \dot{\omega} P_d \left[ \sum_{j=1}^4 \langle \vec{\xi}_j', \vec{\rho} \rangle \vec{e}_j  + \sum_{j=1}^4 \langle \vec{\xi}_j, \vec{\rho} \rangle \vec{e}_j' \right]  \\
%& = - \dot{\omega} \left[ \sum_{j=1}^4 \sum_{k=3}^4 \langle \vec{\xi}_j', \vec{e}_k \rangle\rho_k \vec{e}_j + \rho_3 P_d \vec{e}_3' + \rho_4 P_d \vec{e}_4' \right]  \\
%& =  - \dot{\omega} \left[ \sum_{j=1}^4 \sum_{k=3}^4 \langle \vec{\xi}_j', \vec{e}_k \rangle\rho_k \vec{e}_j + \rho_3 (\Gamma_{13} \vec{e}_1 + \Gamma_{33} \vec{e}_3) + \rho_4 (\Gamma_{24} \vec{e}_2 + \Gamma_{44} \vec{e}_4)   \right] \\
%& = - \dot{\omega} \left[  \langle \vec{\xi}_1', \vec{e}_3 \rangle \rho_3 \vec{e}_1 + \langle \vec{\xi}_2', \vec{e}_4 \rangle \rho_4 \vec{e}_2 + \langle \vec{\xi}_3', \vec{e}_3 \rangle \rho_3 \vec{e}_3 + \langle \vec{\xi}_4', \vec{e}_4 \rangle \rho_4 \vec{e}_4 \right] \\
%&  - \dot{\omega} \left[ \rho_3 (\Gamma_{13} \vec{e}_1 + \Gamma_{33} \vec{e}_3) + \rho_4 (\Gamma_{24} \vec{e}_2 + \Gamma_{44} \vec{e}_4) \right] \\
%& =  - \dot{\omega} \left[ (\langle \vec{\xi}_1', \vec{e}_3 \rangle + \Gamma_{13} ) \rho_3 \vec{e}_1 + (\langle \vec{\xi}_2', \vec{e}_4 \rangle + \Gamma_{24}) \rho_4 \vec{e}_2 \right. \\
%& + \left. (\langle \vec{\xi}_3', \vec{e}_3 \rangle + \Gamma_{33}) \rho_3 \vec{e}_3 + (\langle \vec{\xi}_4', \vec{e}_4 \rangle + \Gamma_{44}) \rho_4 \vec{e}_4  \right].
%\end{align*}

Therefore, the relevant nonzero terms are
\begin{align*}
\langle \vec{\xi}_1, P_d'\vec{\rho} \rangle & =  (\langle \vec{\xi}_1', \vec{e}_3 \rangle + \Gamma_{13})\rho_3, \\
\langle \vec{\xi}_2, P_d'\vec{\rho} \rangle & =  (\langle \vec{\xi}_2', \vec{e}_4\rangle + \Gamma_{24})\rho_4.
\end{align*}

We denote
\begin{align*}
p_{13} & =  \langle \vec{\xi}_1'(\omega^*), \vec{e}_3(\omega^*) \rangle , \\
p_{33} & =  \langle \vec{\xi}_3'(\omega^*), \vec{e}_3(\omega^*) \rangle ,
\end{align*}
and
\begin{align*}
p_{24} &  = \langle \vec{\xi}_2'(\omega^*), \vec{e}_4(\omega^*) \rangle ,\\
p_{44} & = \langle \vec{\xi}_4'(\omega^*), \vec{e}_4(\omega^*) \rangle .
\end{align*}
Note that some cancellation will occur with the $\Gamma_{ij}$ terms that appear separately in the system of ODEs, leaving only these $p_{ij}$ terms in the finally system.

Finally, the terms $\langle \vec{\xi}_i, \vec{N} (\omega, \vec{\rho}) \rangle$ must be computed.   We are only interested in the quadratic terms, which, according to \eqref{quad} are:
\begin{equation}
3Jg'(\phi^2)\phi\rho_{Re}^2 + 2Jg''(\phi^2)\phi^2\rho_{Re}^2+Jg'(\phi^2)\phi\rho_{Im}^2+2g'(\phi^2)\phi\rho_{Re} \rho_{Im}.
\end{equation}
Recall that, since $\rho_1$ and $\rho_2$ are $0$, the projection onto the discrete-spectrum of $\rho_{Re}$ is just $\rho_3 \vec{e}_3$ and the projection onto the discrete-spectrum of $\rho_{Im}$ is just $\rho_4 \vec{e}_4$.  We now have to compute the lowest-order terms of
$$\langle \vec{\xi}_1, \vec{N}(\omega, \vec{\rho}) \rangle.$$

The multiplier of $\rho_3^2$ in $\langle \vec{\xi}_1,\vec{N}(\omega, \vec{\rho}) \rangle$ is
\begin{equation*}
n_{133}=\langle {\xi}_1, (3g'(\phi^2)\phi+ 2g''(\phi^2)\phi^2)e_3^2 \rangle.
\end{equation*}
Similarly, we define
\begin{align*}
n_{144} &= \langle \xi_1, g'(\phi^2)\phi e_4^2 \rangle, \\
n_{234} &= \langle \xi_2, -2g'(\phi^2)\phi e_3e_4 \rangle, \\
n_{333} &= \langle \xi_3, (3g'(\phi^2)\phi+ 2g''(\phi^2)\phi^2)e_3^2 \rangle,  \\
n_{344} &= \langle \xi_3, g'(\phi^2)\phi e_4^2 \rangle, \\
n_{434} &= \langle \xi_4, -2g'(\phi^2)\phi e_3e_4 \rangle.
\end{align*}
Notice that, as in the computation of the $c_{ij}$, these are real inner products between real functions that normally appear in different components of the complex vectors.

Lastly, we need to estimate $a(\omega)$.  Recall that $a(\omega^*)=0$, and that $a(\omega)$ appears in \eqref{fullsystem1} multiplied by $\rho_3$, so we are seeking only the linear term, $a(\omega) \sim a_0 (\omega - \omega^*)$.  We calculate:
\begin{eqnarray*}
a_0 = a' (\omega^*) = -\frac{2}{\langle \phi_{\omega^*} , \beta \rangle} (\langle \phi_{\omega^*}', \phi_{\omega^*}' \rangle - \langle \phi_{\omega^*} , \phi_{\omega^*}'' \rangle).
\end{eqnarray*}

With these assumptions, we conclude the following:

\begin{prop}  The quadratic approximation for the evolution of a perturbation of the minimal mass soliton, \eqref{ansatzequation}, ignoring coupling to the continuous spectrum, is
\begin{align}
\dot{\theta} & =- c_{14}\dot{\theta}\rho_4 +n_{133}\rho_3^2+n_{144}\rho_4^2 + \dot{\omega} p_{13} \rho_3, \notag  \\
\dot{\omega} & = \rho_3 + c_{23}\dot{\theta} \rho_3+n_{234}\rho_3\rho_4 +\dot{\omega} p_{24} \rho_4, \notag \\
\dot{\rho_3} & = \rho_4 - c_{34}\dot{\theta}\rho_4+n_{333}\rho_3^2+n_{344}\rho_4^2 +\dot{\omega} p_{33} \rho_3, \label{quadraticsystem} \\
\dot{\rho_4} & = a_0 (\omega - \omega^*) \rho_3 + c_{43}\dot{\theta}\rho_3+n_{434}\rho_3\rho_4 +\dot{\omega} p_{44}\rho_4. \notag
\end{align}
\end{prop}
In this system we implicitly assume that $\theta$,
$(\omega-\omega^*)$, $\rho_i$ and their time derivatives are all of the same order.

\section{Numerical Methods}

From here on, we use numerical techniques to analyze solutions to \eqref{quadraticsystem}.  We will work in one space dimension, with the specific saturated nonlinearity $g(s)=\frac{s^3}{1+s^2}$ as described in the introduction.     Though we have a complete description of the generalized kernel of $J\mathcal{H}$, including its size and the relation among the elements, nothing is expressible in terms of elementary functions.  As this kernel  determines the coefficients in our ODE system, we numerically compute it, permitting us to subsequently integrate the ODEs numerically.

The sinc function, $\sin(\pi x) / (\pi x)$ was used to compute solitary wave solutions when analytical expressions were not readily available in \cite{lundin1980cfm} and in the forthcoming \cite{simpson09}.  It has also been used to study time dependent nonlinear wave equations, \cite {alkhaled2001sns,revelli2003sci, bellomo1995sni, carlson1997scm},  and a variety of linear and nonlinear boundary value problems, \cite{bialecki1989sta,bialecki1991scm,elgamel2003nms,elgamel2004sgm,elgamel2007san,mohsen2008gac}.

We will use the sinc function to estimate the coefficients in three steps:
\begin{itemize}
\item Compute a discrete representation of the minimal mass soliton, $\phi_{\omega^\ast}$.
\item Compute discrete representations of the generalized kernel of $\mathcal{H}$, i.e. the derivatives with respect to $\omega$.
\item Compute necessary inner products for the coefficients.
\end{itemize}

\subsection{Sinc Discretization}
\label{sec:sinc_disc}
The problem of finding a soliton solution of \eqref{eqn:sol} is a nonlinear boundary value problem posed on $\mathbb{R}$.  We respect this description in our discretization by approximating functions with the $\sinc$ spectral method.  This technique is thoroughly explained in \cite{lund1992smq, stenger1993nmb, stenger1981nmb, stenger2000ssn} and briefly in Appendix \ref{app:sinc_details}.  In the $\sinc$ discretization, the problem remains posed on $\mathbb{R}$ and the boundary conditions, that the solution vanish at $\pm \infty$, are naturally incorporated.

Given a function $u(x):\mathbb{R} \to \mathbb{R}$, $u$ is approximated using a superposition of shifted and scaled $\sinc$ functions:
\begin{equation}
\label{eq:sinc_approx}
C_{M, N}(u,h)(x) \equiv \sum_{k = -M}^N u_k \sinc\paren{\frac{x-x_k}{h}} = \sum_{k = -M}^N u_k S(k,h) (x) ,
\end{equation}
where $x_k = k h$ for $k = -M, \ldots, N$ are the \emph{nodes} and $h>0$.  There are \emph{three} parameters in this discretization, $h$, $M$, and $N$, determining the number of and spacing of lattice points.  This is common to numerical methods posed on unbounded domains; see \cite{boyd2001caf}.

A useful and important feature of this spectral method is that, when evaluated at a node,
\begin{equation}
\label{eq:disc_delta}
C_{M,N}(u,h)(x_k) = u_k .
\end{equation}
Additionally, the convergence is rapid both in practice and theoretically.  See Theorem \ref{thm:sinc_convergence} in Appendix \ref{app:sinc_details} for a statement on optimal convergence.

Since the soliton is an even function, we may take $N = M$.  We will thus write
\begin{equation}
C_{M}(u,h)(x) \equiv C_{M,M}(u,h)(x) .
\end{equation}
The symmetry implies $u_{-k} = u_{k}$ for $k = -M,\ldots M$.  We take advantage of this constraint in our computations.  In addition, we slave $h$ to $M$ in accordance with \eqref{eq:h_prag}.

To compute a discrete sinc approximation of the ground state, we frame the soliton equation as a \emph{nonlinear collocation problem.}  Approximating $\phi(x)$ as in \eqref{eq:sinc_approx}, we seek coefficients $\set{R_k}$ such that
\begin{equation}
\label{eq:collocation_eqs}
\begin{split}
&\partial_x^2 C _M(\phi, h)(x_k) - \lambda  C _M(\phi, h)(x_k)\\
&\quad  + g(\abs{ C _M(\phi, h)(x_k) }^2) C _M(\phi, h)(x_k) =0, \quad\text{for $k = -M,\ldots, M$} .
\end{split}
\end{equation}
By satisfying \eqref{eq:collocation_eqs}, the discrete approximation solves the soliton equation in the \emph{strong} sense at the nodes, also known as \emph{collocation points}.  This is in contrast to a Galerkin formulation, which solves the equation in the weak sense.  However, for the one dimensional under consideration, sinc-Galerkin and sinc-collocation lead to the same algebraic system.

\eqref{eq:collocation_eqs} yields a system of nonlinear algebraic equations.  Let $\vec{\phi}$ be the column vector associated with the discrete approximation of $\phi$:
\begin{equation}
C_M(\phi,h)(x_k) \mapsto \vec{\phi} = \begin{pmatrix} \phi_{-M}\\ \phi_{-M+1}\\\vdots\\ \phi_M \end{pmatrix} .
\end{equation}
Differentiation of a $\sinc$ approximated function that is evaluated at the collocation points corresponds to matrix multiplication:
\begin{equation}
\partial_x^2 C_M(\phi,h)(x_k) \mapsto D^{(2)} \vec{\phi} .
\end{equation}
Explicitly, $D^{(2)}$ is
\begin{equation}
\label{eq:matrix_d2}
D^{(2)}_{jk} = \frac{d^2}{dx^2}S(j,h)(x)|_{x = x_k} = \begin{cases} \frac{1}{h^2} \frac{-\pi^2}{3} & j=k \\ \frac{1}{h^2} \frac{-2 (-1)^{k-j}}{(k-j)^2} & j \neq k \end{cases} .
\end{equation}

Using \ref{eq:disc_delta},
\[
g(\abs{C_M(\phi,h)(x_k)}^2)C_M(\phi,h)(x_k) = C_M(g(\abs{\phi}^2) \phi, h)(x_k)= g(\abs{\phi_k}^2) \phi_k .
\]
Thus
\[
g(\abs{C_M(\phi,h)(x_k)}^2)C_M(\phi,h)(x_k)\mapsto g(|{\vec{\phi}}|^2) \vec{\phi} \equiv \begin{pmatrix} g(\abs{\phi_{-M}}^2) \phi_{-M}\\ \vdots\\ g(\abs{\phi_M}^2) \phi_M \end{pmatrix} .
\]
With these relations, the discrete system is
\begin{equation}
\label{eq:disc_solition_eq}
D^{(2)} \vec{\phi} - \omega \vec{\phi} +  g(|\vec{\phi}|^2) \vec{\phi}=0 .
\end{equation}
It is this equation to which we apply a nonlinear solver, subject to an appropriate guess.  We discuss an important subtlety in Appendix \ref{sec:continuation}.

\subsection{Computing the Minimal Mass}
Now that we have an algorithm for finding a discrete representation of a soliton, we seek to find the value of the soliton parameter for the one possessing minimal mass, along with the corresponding discretized soliton.  The $\sinc$ discretization has the property that the $L^2(\mathbb{R})$ inner product is well approximated by
\[
\inner{f}{g} = \int f g dx \approx \sum_{-M}^M h f_k g_k = h (\vec{f} \cdot \vec{g}) .
\]
Thus, the mass of the soliton can be estimated by
\[
\norm{\phi}_{L^2}^2 \approx h |{\vec{\phi}}|^2 .
\]
Recognizing that $\vec{\phi} = \vec{\phi}(\omega)$, we seek to minimize the functional $h |{\vec{\phi}(\omega)}|^2$ with respect to $\omega$.  The argument $\omega$ for which the minimum occurs will be $\omega^\ast$.  To find the minimal mass, we take the derivative, getting a discrete representation of the minimal mass orthogonality condition:
\begin{equation}
\label{eq:disc_minmass_ortho}
2 h \vec{\phi} \cdot \vec{\phi'} = 0 .
\end{equation}
We solve \eqref{eq:disc_minmass_ortho} to find $\omega^\ast$, computing $\vec{\phi_{\omega^\ast}}$ in the process.

The value $\omega^\ast$ can be obtained by other algorithms.  In the one-dimensional case, the soliton equation possesses a first integral, permitting the minimal mass to be computed by numerical quadrature and minimization; a comparison of our results and this approach appears in Appendix \ref{sec:quad_root}.  Though these approaches are quite accurate for the task of computing the minimal mass, they are inadequate at computing the generalized kernel.  Thus, we seek to solve the problem consistently by finding the minimal mass for a given $2M+1$ dimensional approximation of the problem.

\subsection{Discretized Generalized Kernel}
Formally, at the minimal mass soliton $\phi_{\omega^\ast}$, there are four functions associated with the kernel satisfying the second order equations:
\begin{align}
L_- \phi_{\omega^\ast} &= 0 ,\\
L_+ (-\phi_{\omega^\ast}') & = \phi_{\omega^\ast} ,\\
L_- \alpha & = -\phi_{\omega^\ast}' ,\\
L_+ \beta & = \alpha .
\end{align}
These four functions can also be discretized with $\sinc$, as in \eqref{eq:sinc_approx}.  The operators, $L_\pm$, have discrete spectral representations:
\begin{align}
\label{eq:disc_Lp}
L_+ &\mapsto \mathbf{L}_+\equiv -D^{(2)} + \omega I  - \diag\{g(\vec{\phi}_{\omega^\ast}^2)\} ,\\
\label{eq:disc_Lm}
L_- &\mapsto \mathbf{L}_-\equiv-D^{(2)} + \omega I  - \diag\{g(\vec{\phi}_{\omega^\ast}^2) -2 g' ( \vec{\phi}_{\omega^\ast}^2) \vec{\phi}_{\omega^\ast}^2\} .
\end{align}
Taking $\vec{u} = \vec{\phi}$, we successively solve for $\vec{\phi'}$, $\vec{\alpha}$, and $\vec{\beta}$.  A singular value decomposition must be used to get $\vec{\alpha}$ since  $\mathbf L _ -$ has a non-trivial kernel.

Furthermore, we compute discrete approximations of the derivatives of $\phi$, $\phi'$, $\alpha$, and $\beta$ taken with respect to $\omega$ at $\omega^\ast$.  The relevant operators are formed analogously to \eqref{eq:disc_Lp} and \eqref{eq:disc_Lm}.

\subsection{Convergence}
Amongst the many calculations made, the most important is of $\omega^\ast$, the  parameter of the minimal mass soliton.  We summarize the results in Table \ref{tab:mass_conv}.   We see that $h |\vec{\phi}|^2$ robustly converges, achieving twelve digits of precision and  $\omega^\ast$ appears to achieve eleven digits of precision.  These are consistent with the values in Table \ref{table:quadmin_results} from Appendix \ref{sec:quad_root}, where they were computed using a different methods.

For the purposes of our simulations, we believe we have sufficient precision, approximately ten significant digits, for the time integration of our system of ODEs.  Some data for the convergence of the coefficients appearing in \eqref{quadraticsystem} is given in Appendix \ref{sec:convergence_data}.

\begin{table}
\caption{The convergence of the sinc discretization to the minimal mass soliton.}
\label{tab:mass_conv}
\begin{tabular}{r|l|l}
 $M$ & $h |\vec{\phi}|^2$ & $\omega^\ast$ \\
 \hline
  20 & 3.820771417633398 & 0.177000229690401 \\
  40 & 3.821145471868853 & 0.177576993694258 \\
  60 & 3.821148930202135 & 0.177587655985074 \\
  80 & 3.821149018422933 & 0.177588043323139 \\
 100 & 3.821149022493814 & 0.177588063805561 \\
 200 & 3.821149022780618 & 0.177588065432740 \\
 300 & 3.821149022780439 & 0.177588065432795 \\
 400 & 3.821149022780896 & 0.177588065433095 \\
 500 & 3.821149022780275 & 0.177588065432928
 \end{tabular}
\end{table}

\section{Numerical Results}
\label{nm}

We explore here the dynamics of the finite dimensional system \eqref{quadraticsystem} and compare with solutions for the full nonlinear PDE \eqref{nls} with corresponding initial data.

To solve \eqref{quadraticsystem}, we use the stiff solver {\it ode45} from {\sc Matlab} after properly preparing the initial data using the soliton finding codes in Section \ref{sec:sinc_disc}.

\subsection{Nonlinear Solver}
In order to determine the accuracy of our results, we also use a nonlinear solver to approximate the solutions with a perturbed minimal-mass soliton as initial data.  For this nonlinear solver, we use a finite element scheme in space and a Crank-Nicholson scheme in time.  This is similar to the method used in \cite{HMZ}.  In brief, we discretize our \eqref{nls} by method of lines, using finite elements in space and Crank-Nicholson for time-stepping. This method is $L^2$ conservative, though it is not energy conserving.  A similar scheme was implemented without potential in {\cite{ADKM3}}, where
the blow-up for NLS in several dimensions was investigated.
%\begin{eqnarray*}
%\left \{ \begin{array}{c}
%i u_t + u_{xx} + g(|u|^2) u = 0, \\
%u(0,x) = u_0,
%\end{array} \right.
%\end{eqnarray*}

We require the spatial grid to be large enough to ensure negligible interaction with the boundary.
As absorbing boundary conditions for cubic NLS currently require high frequency limits to apply successfully, we choose simply to carefully ensure that our grid is large enough in order for the interactions to be negligible throughout the experiment. For the convergence of such methods without potentials see the references in \cite{ADKM3}, {\cite{ADKM1}} and {\cite{ADKM2}}.

We select a symmetric region about the origin, $[-R,R]$, upon which
we place a mesh of $N$ elements.  The standard hat function basis is
used in the Galerkin approximation. We allow for a finer grid in a
neighbourhood of length $1$ centered at the origin to better study the effects of the soliton interactions.

%One can rescale the problem to have a grid from $[-1,1]$, however, the same number of grid points are required to accurately solve the problem as the scaling introduces large constants.

In terms of the hat basis the PDE \eqref{nls} becomes:
\begin{align*}
&\langle u_t, v \rangle + {i} \langle u_x, v_x \rangle /2
- i \langle g(|u|^2) u,v \rangle = 0, \\
&u(0,x) = u_0 \,, \ \  u ( t , x) = \sum_{v} c_v ( t) v \,,
\end{align*}
where $\langle \cdot, \cdot \rangle$ is the standard $L^2$ inner product, $v$ is a basis function and $u$, $u_0$ are linear combinations of the $v$'s.

Since the $v$'s are hat functions, we have a tridiagonal linear system.
Let $h_t > 0 $ be a uniform time step, and let
\[ u_n = \sum_{v} c_v ( n h_t ) v  \]
be the approximate solution at the $n$th time step.
Implementing Crank-Nicholson, the system becomes:
\[ \begin{split}
& \langle u_{n+1} - u_n , v\rangle + i h_t \left\langle \left( {(u_{n+1} + u_{n})}/{2} \right)_x, v_x \right\rangle   \\
& \ \ \ \ = i h_t \left\langle g( |({u_{n+1} + u_{n}})/{2}|^2 )({u_{n+1} + u_{n}})/{2} , v \right\rangle, \ \ \ \
 u_0 = \sum_v \alpha_v v .
\end{split} \]
By defining
\[ y_{n} = (u_{n+1} + u_n)/2 \,, \]
we have simplified our system to:
\[
\langle y_n , v \rangle + i \frac{h_t}{4} \langle ({y_{n}})_x,v_{x} \rangle  =
i \frac{h_t}{2} \langle |y_n|^2 y_n, v \rangle  +  \langle u_n, v\rangle.
\]
An iteration method from {\cite{ADKM3}} is now
used to solve this nonlinear system of equations.
Namely, we set,
\[
\langle y^{k+1}_{n} , v \rangle +
i \frac{h_t}{4} \langle (y^{k+1}_n)_x, v_{x} \rangle =
i \frac{h_t}{2} \langle |y_n^k|^2 y_n^k, v \rangle  +  \langle u_n, v \rangle.
\]
We take $y_n^0 = u_n$
and perform three iterations in order to obtain an approximate solution.

For our problem, we have taken \eqref{nls} with the nonlinearity
\begin{equation*}
\frac{|u|^6}{1+|u|^4} u.
\end{equation*}
Then, the minimal mass soliton occurs at
\begin{equation*}
\omega^\ast = .177588065433.
\end{equation*}

\subsection{Results}

\begin{figure}
\includegraphics[width=3.0in]{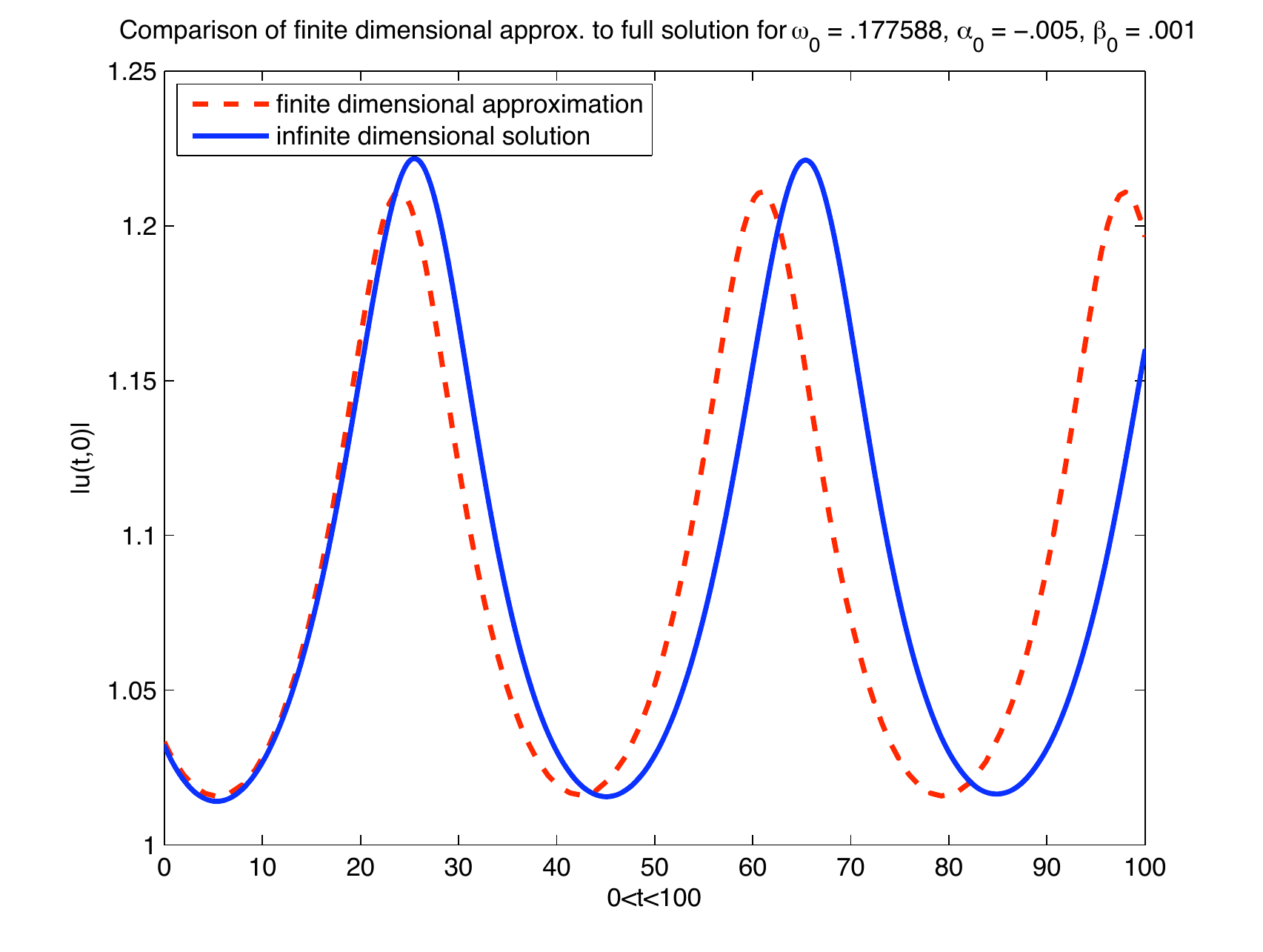}
\includegraphics[width=3.0in]{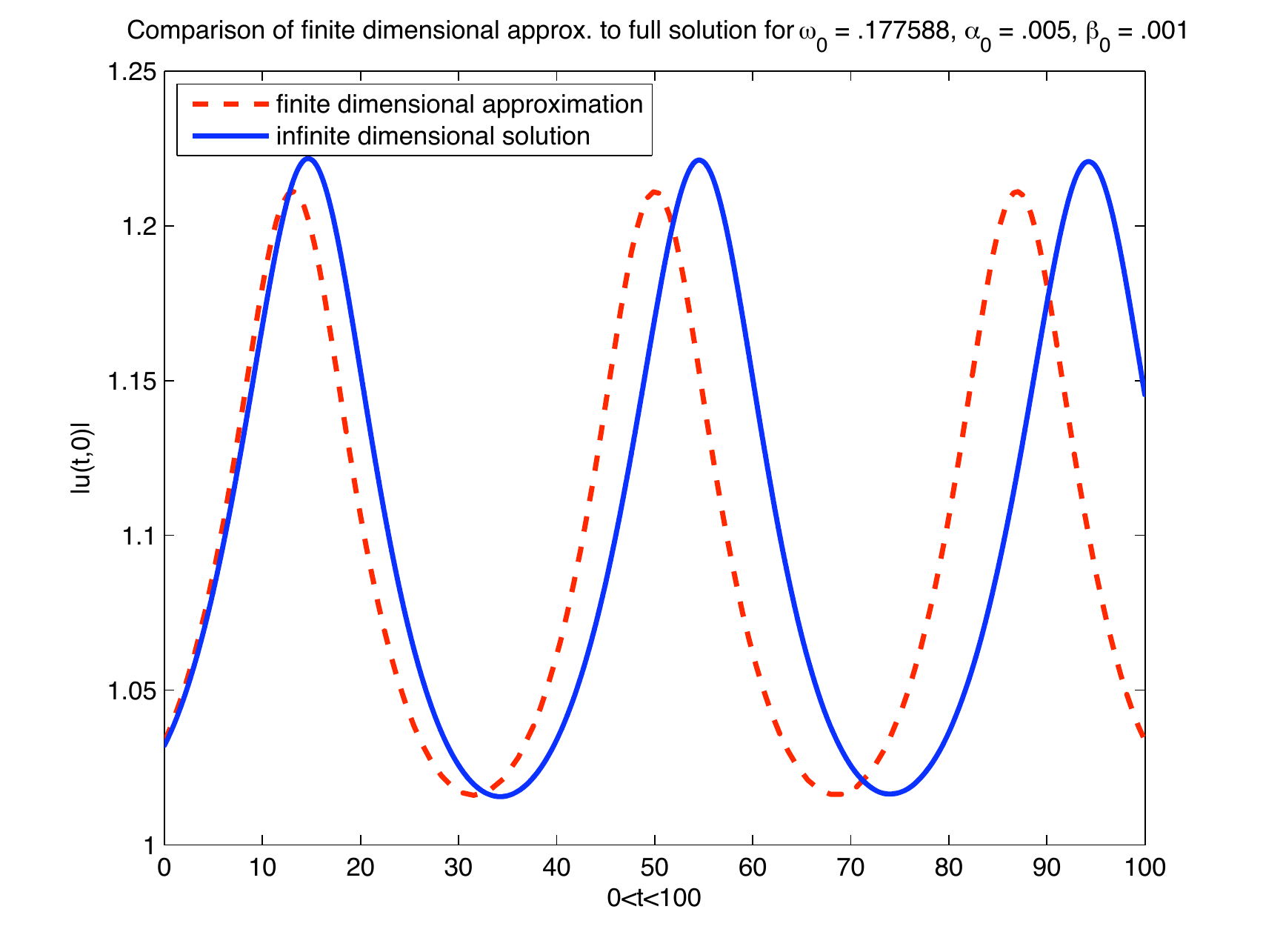}
\caption{A plot of the solution to the system of ODE's as well as the full solution to \eqref{nls} derived for solutions near the minimal soliton for $\rho_3 (0) > 0$ and $\rho_3 (0) < 0$, $\rho_4 (0) > 0$ for $\omega_0 = .177588$, $N = 1000$.}
\label{fig:idfdosc1}
\end{figure}

\begin{figure}
\includegraphics[width=3.0in]{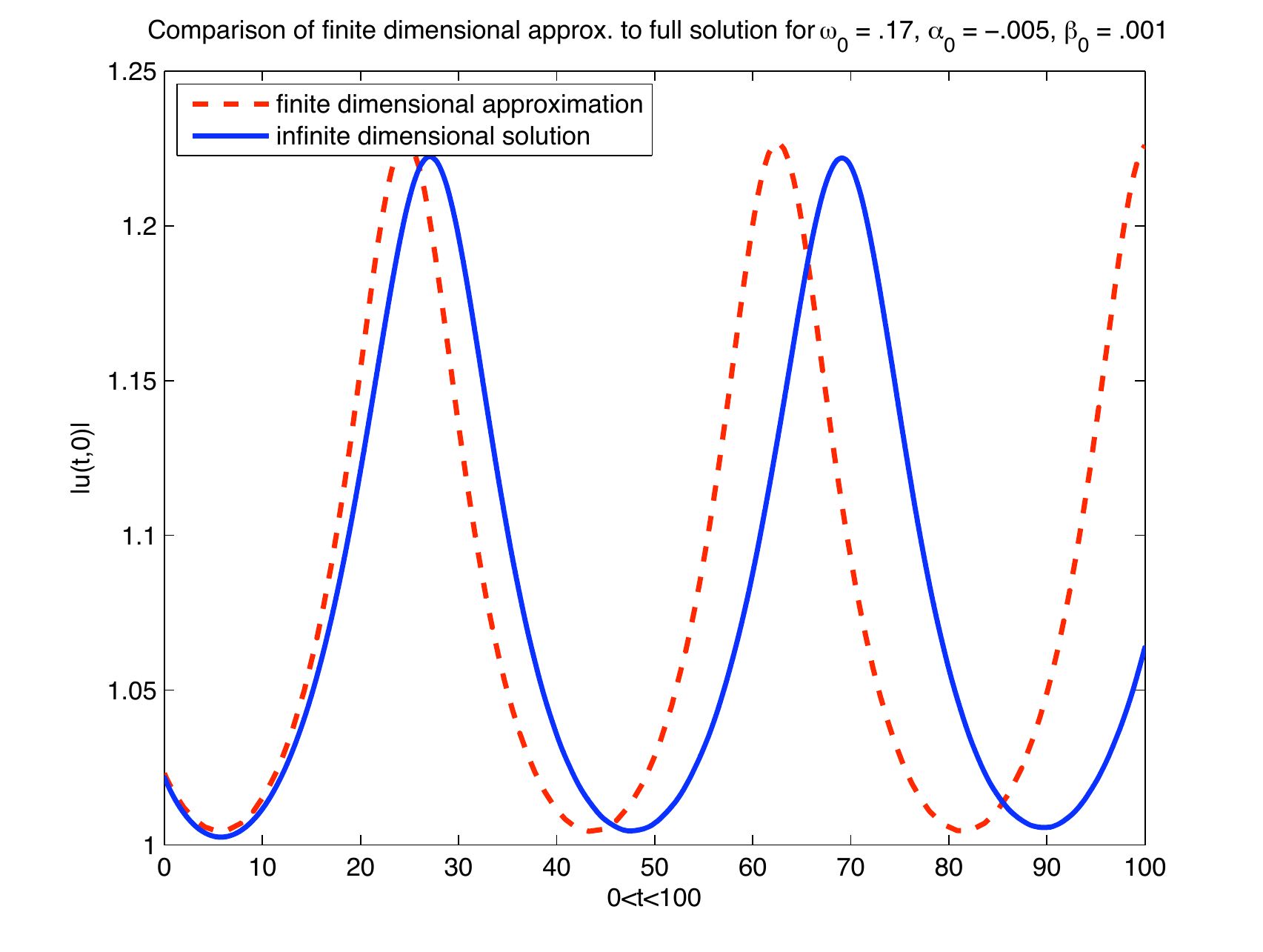}
\includegraphics[width=3.0in]{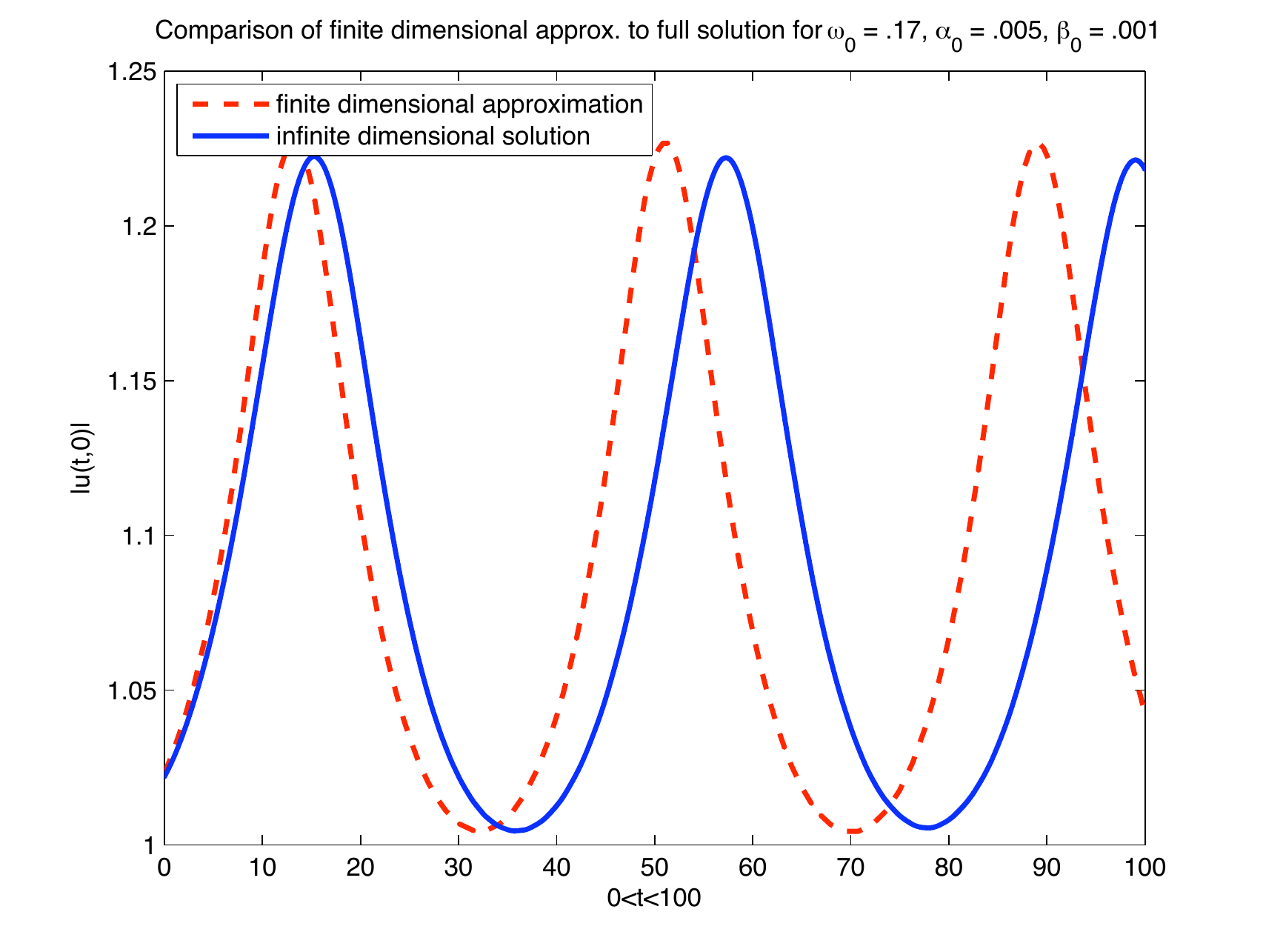}
\caption{A plot of the solution to the system of ODE's as well as the full solution to \eqref{nls} derived for solutions near the minimal soliton for $\rho_3 (0) > 0$ and $\rho_3 (0) < 0$, $\rho_4 (0) > 0$ for $\omega_0 = .17$, $N = 500$.}
\label{fig:idfdosc2}
\end{figure}

\begin{figure}
\includegraphics[width=3.0in]{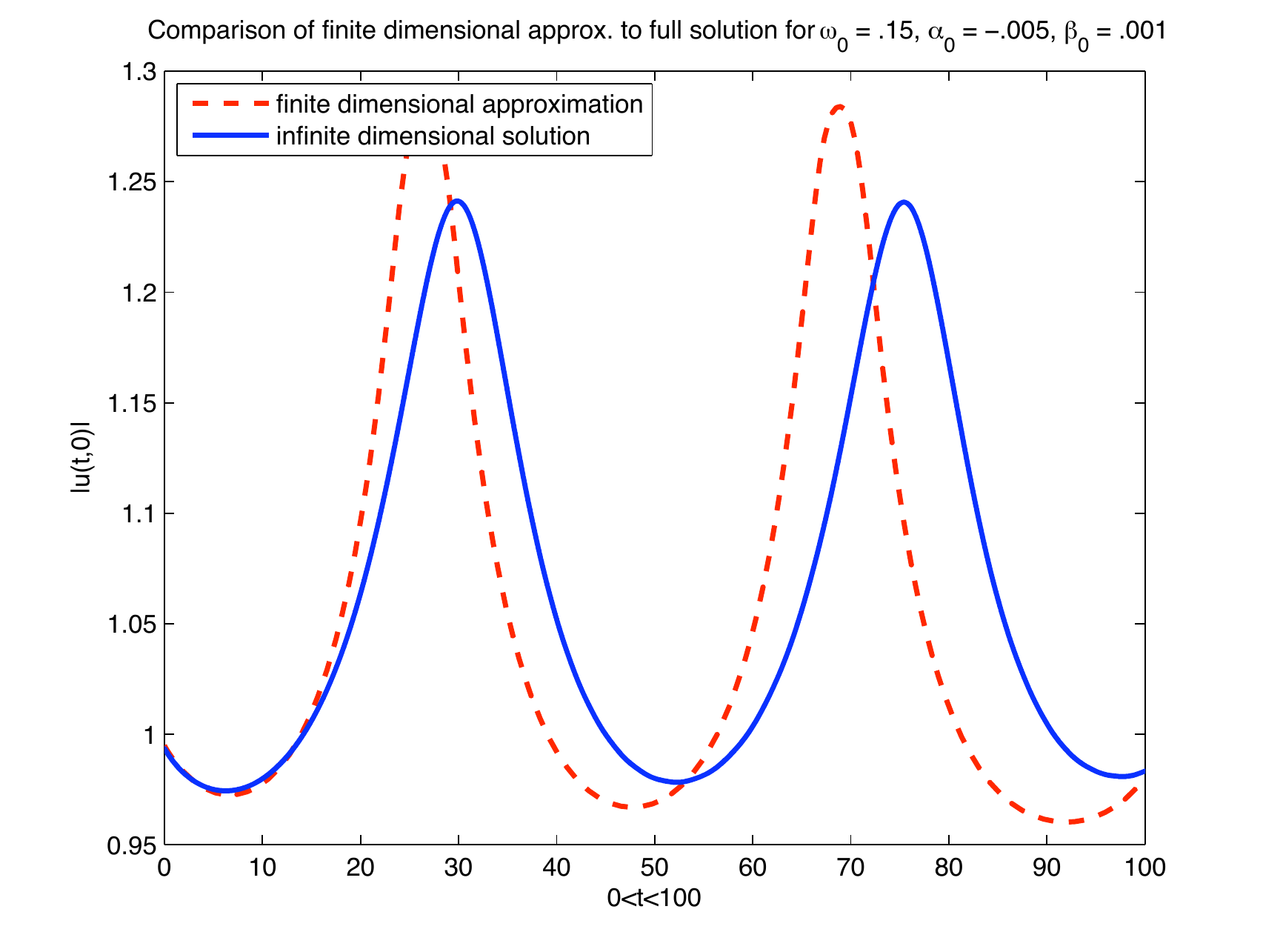}
\includegraphics[width=3.0in]{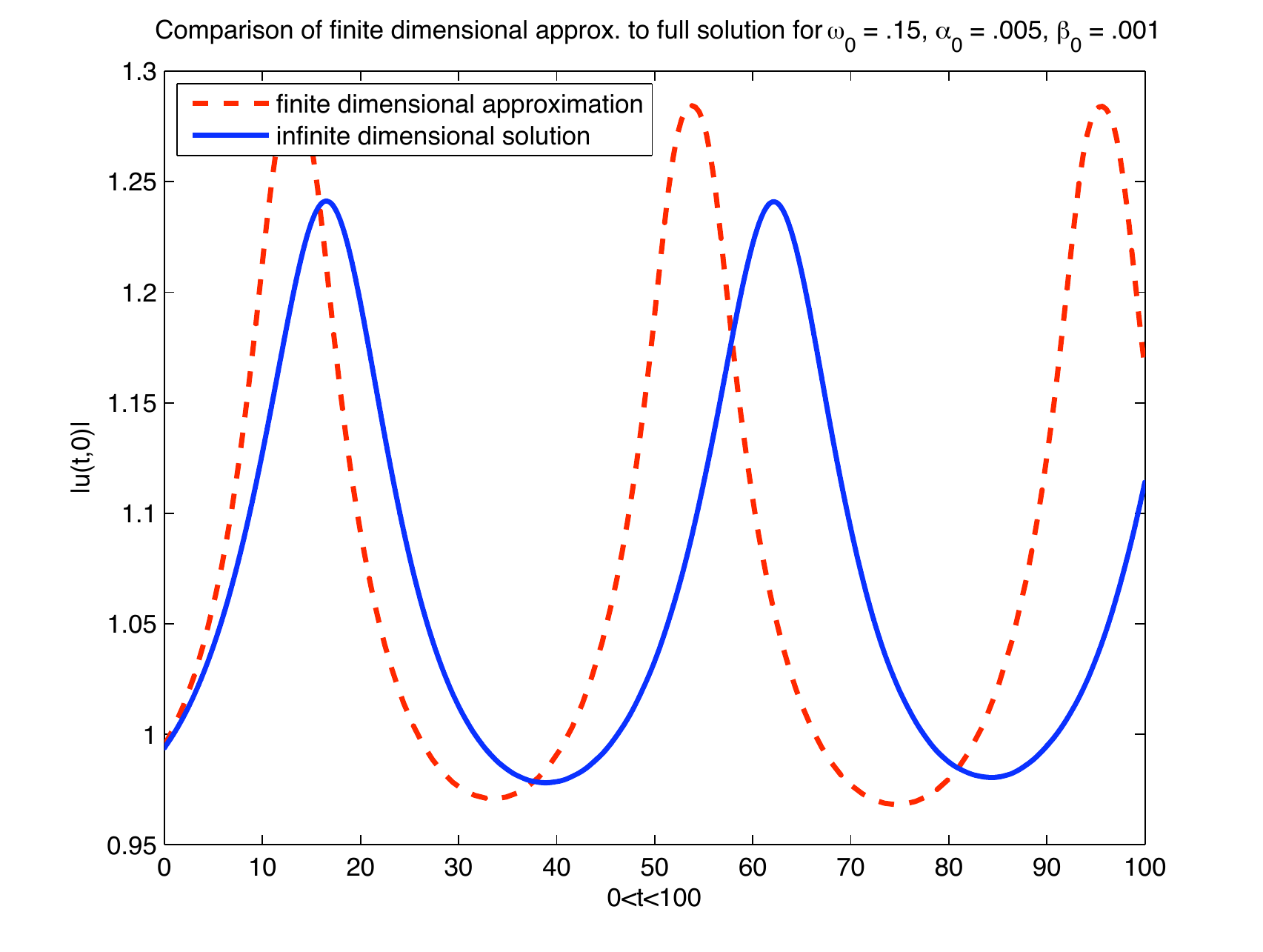}
\caption{A plot of the solution to the system of ODE's as well as the full solution to \eqref{nls} derived for solutions near the minimal soliton for $\alpha_0 = \rho_3 (0) > 0$ and $\alpha_0 = \rho_3 (0) < 0$, $\beta_0 = \rho_4 (0) > 0$ for $\omega_0 = .15$, $N = 500$.}
\label{fig:idfdosc3}
\end{figure}

\begin{figure}
\includegraphics[width=3.0in]{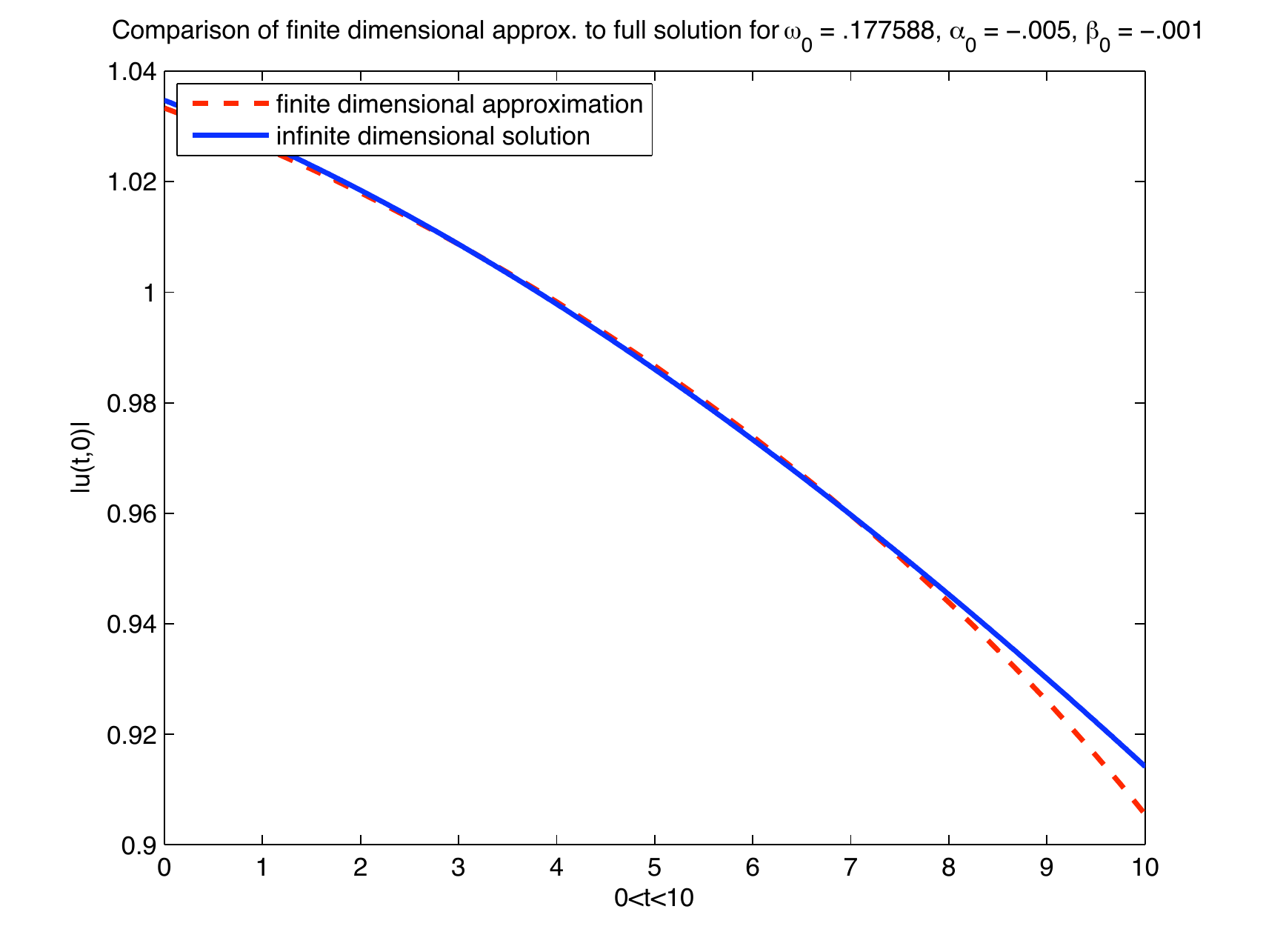}
\includegraphics[width=3.0in]{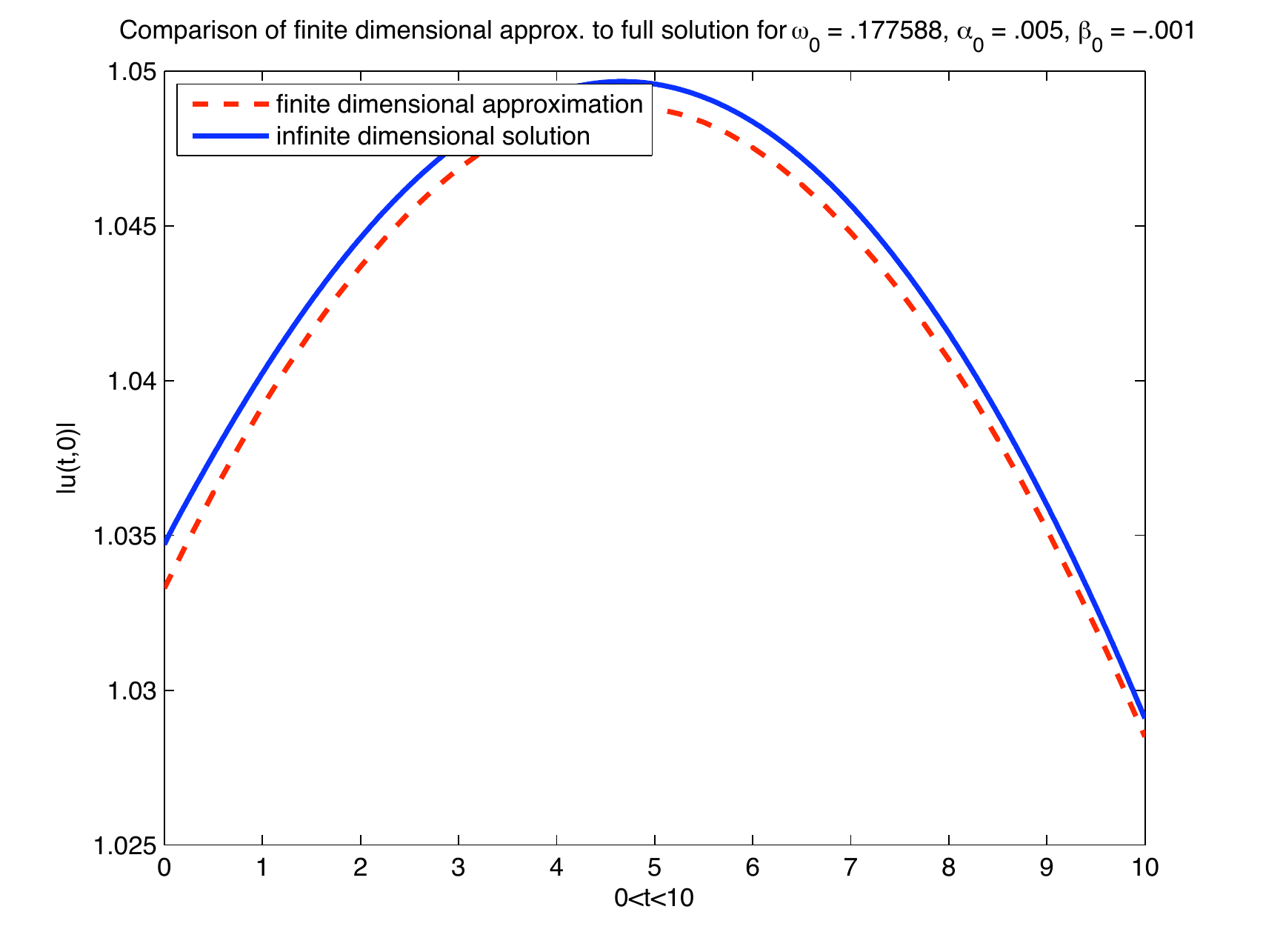}
\caption{A plot of the solution to the system of ODE's as well as the full solution to \eqref{nls} derived for solutions near the minimal soliton for $\alpha_0 = \rho_3 (0) > 0$ and $\alpha_0 = \rho_3 (0) < 0$, $\beta_0 = \rho_4 (0) < 0$ for $\omega_0 = .177588$, $N = 1000$.}
\label{fig:idfdosc4}
\end{figure}

\begin{figure}
\includegraphics[width=3.0in]{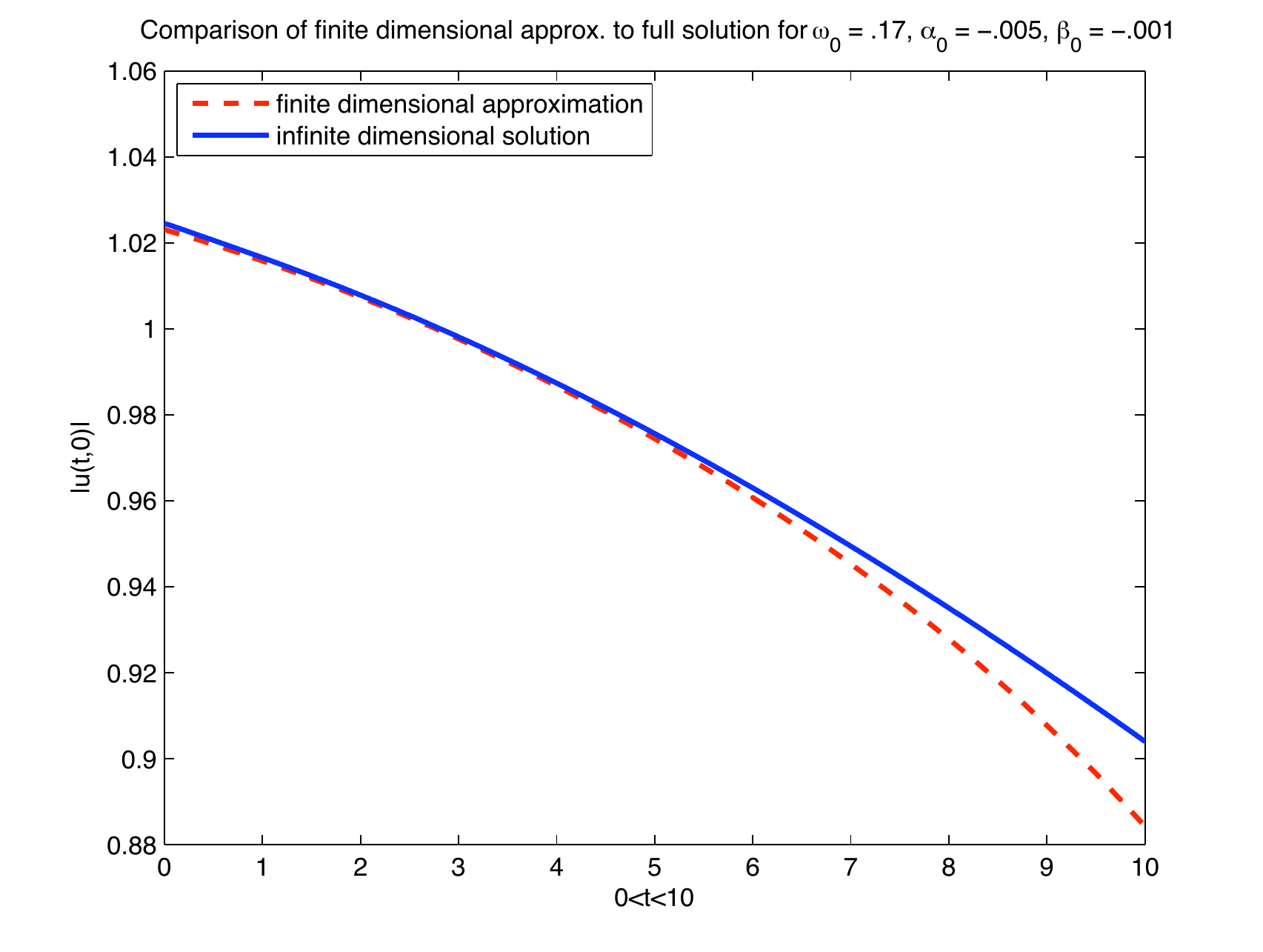}
\includegraphics[width=3.0in]{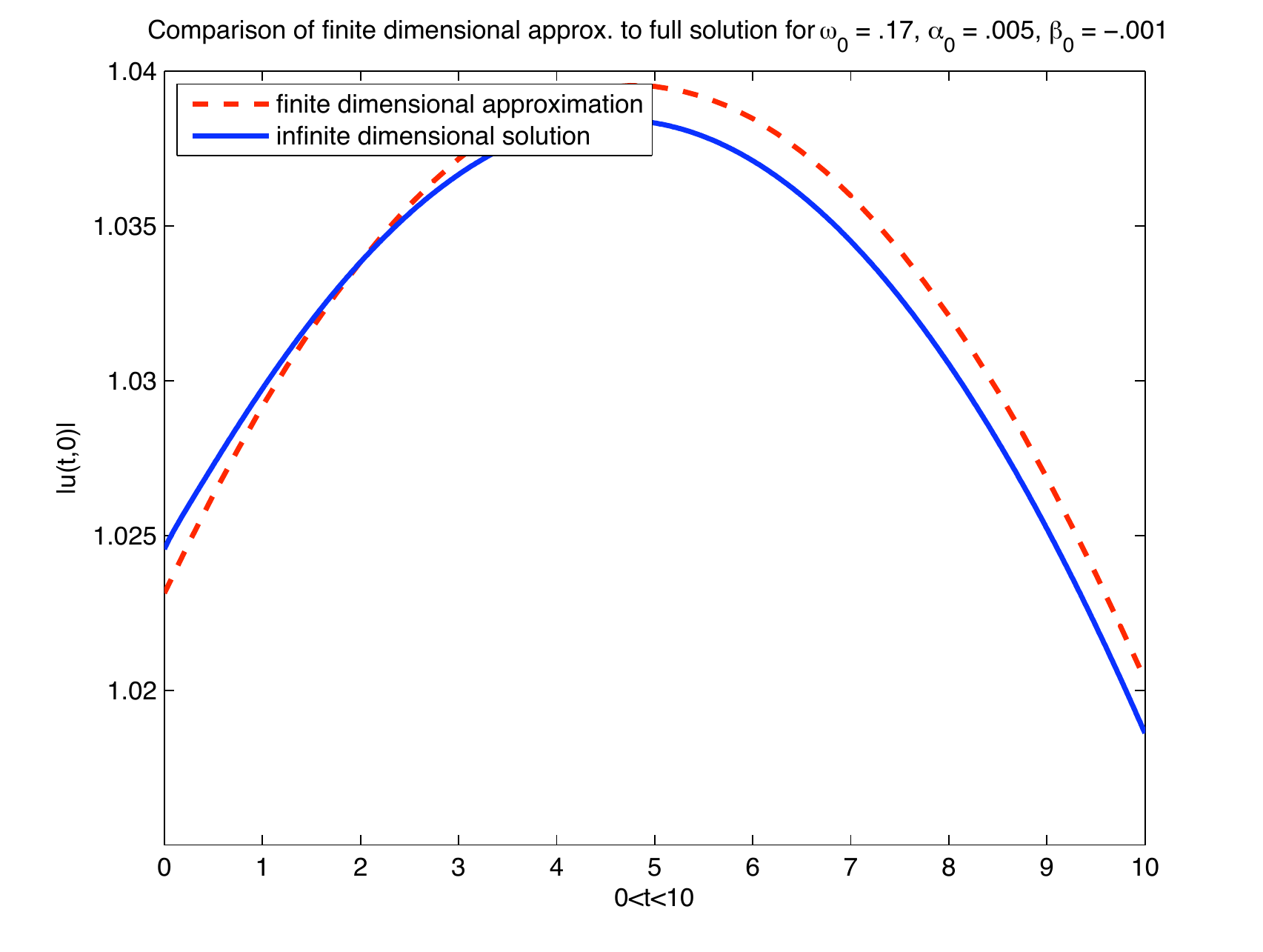}
\caption{A plot of the solution to the system of ODE's as well as the full solution to \eqref{nls} derived for solutions near the minimal soliton for $\alpha_0 = \rho_3 (0) > 0$ and $\alpha_0 = \rho_3 (0) < 0$, $\beta_0 = \rho_4 (0) < 0$ for $\omega_0 = .17$, $N = 500$.}
\label{fig:idfdosc5}
\end{figure}

\begin{figure}
\includegraphics[width=3.0in]{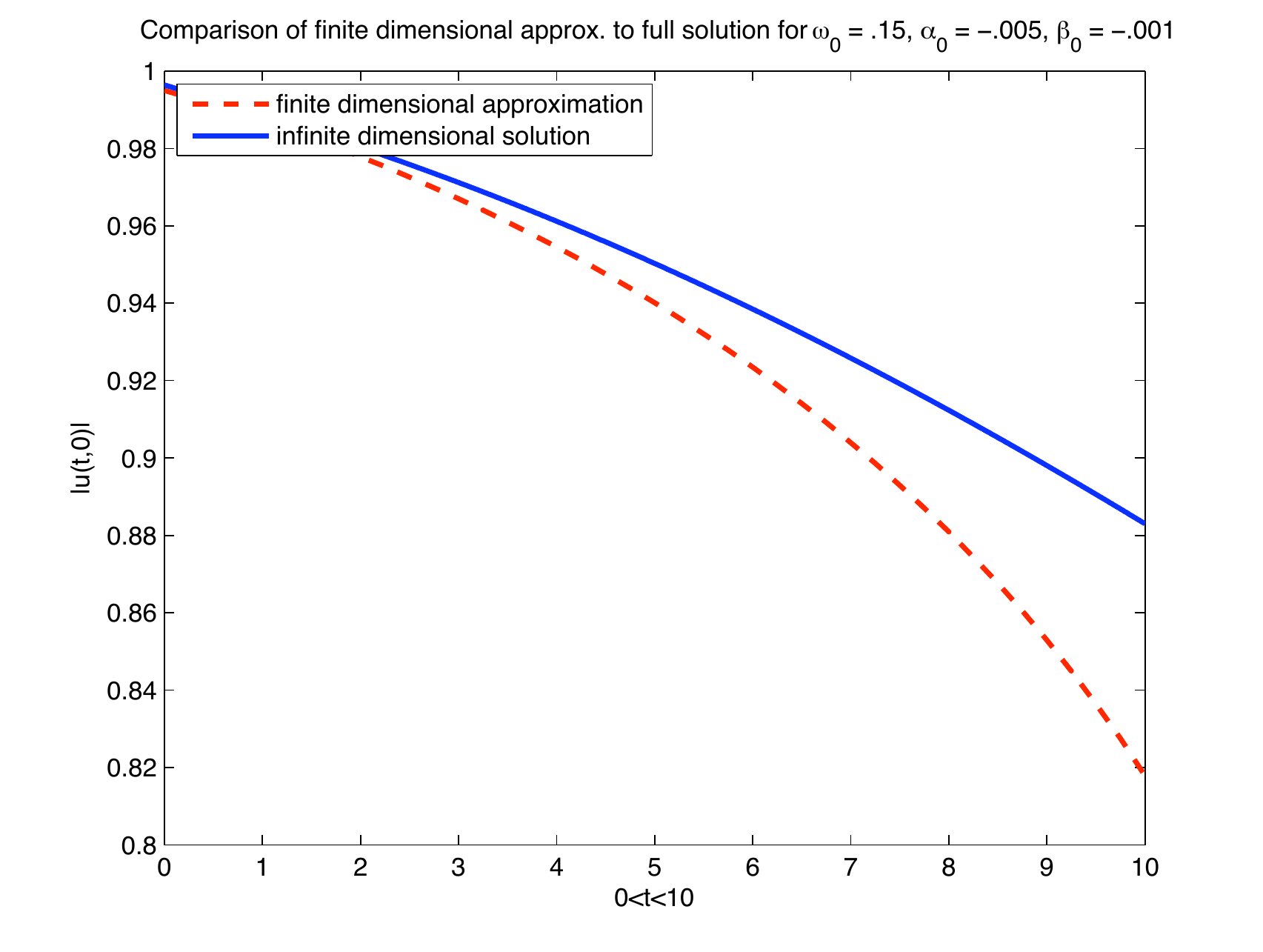}
\includegraphics[width=3.0in]{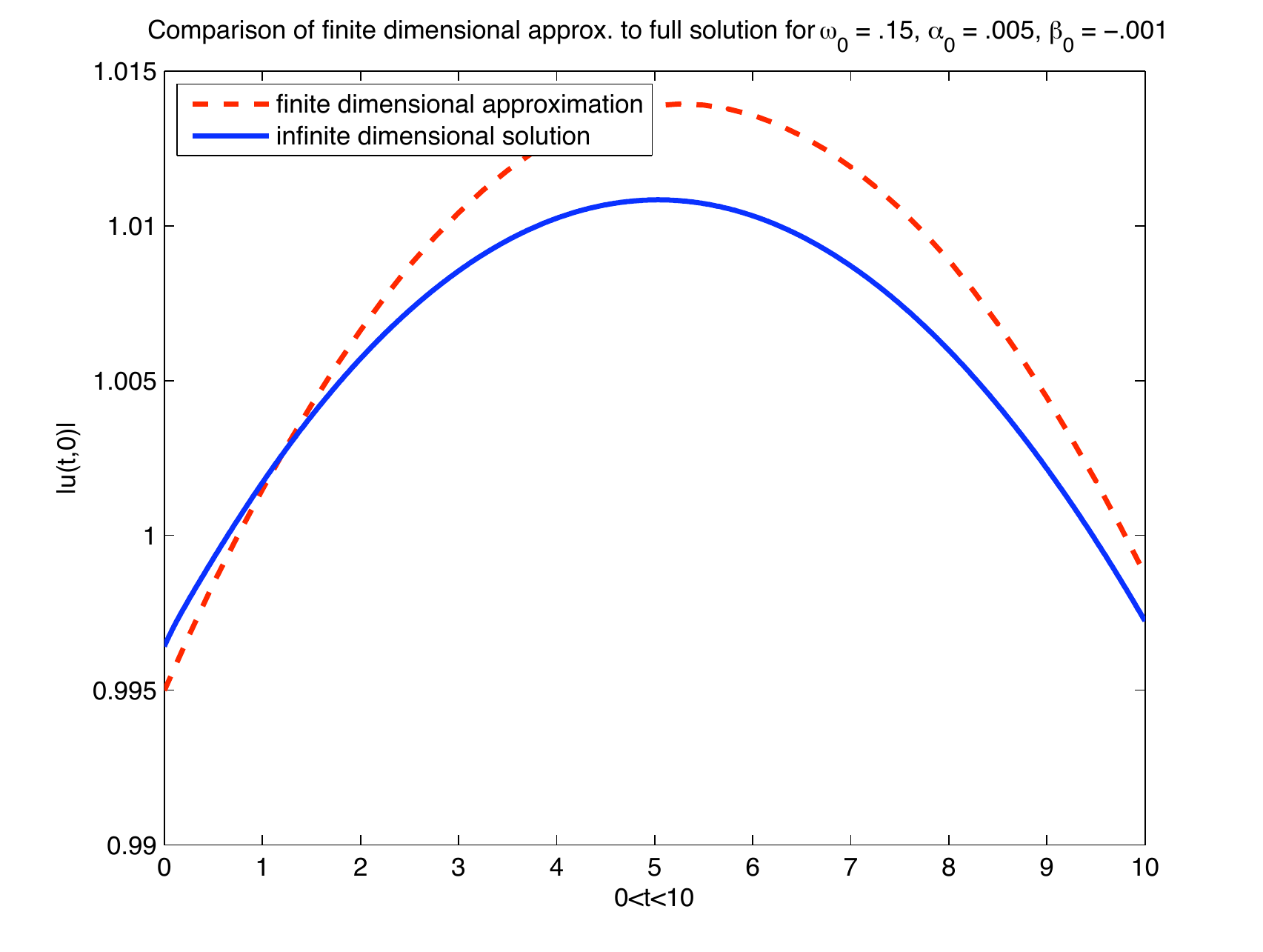}
\caption{A plot of the solution to the system of ODE's as well as the full solution to \eqref{nls} derived for solutions near the minimal soliton for $\alpha_0 = \rho_3 (0) > 0$ and $\alpha_0 = \rho_3 (0) < 0$, $\beta_0 = \rho_4 (0) < 0$ for $\omega_0 = .15$, $N = 500$.}
\label{fig:idfdosc6}
\end{figure}

With the numerical schemes outlined above, we then compare our finite dimensional model to the numerically integrated solution with appropriate initial data $\omega_0 = \omega (0)$, $\alpha_0 = \rho_3 (0)$, $\beta_0 = \rho_4 (0)$ and  $\theta_0 = \theta (0) = 0$ for simplicity.  In Figures \ref{fig:idfdosc1}, \ref{fig:idfdosc2}, \ref{fig:idfdosc3}, we take $\beta_0 > 0$ and vary $\alpha_0$, $\omega_0$.  Similarly, in Figures \ref{fig:idfdosc4}, \ref{fig:idfdosc5}, \ref{fig:idfdosc6}, we take $\beta_0 < 0$ and once again vary $\alpha_0$, $\omega_0$.  Note that we are comparing solutions to the ODEs to solutions of \eqref{nls} with the correct initial parameters so that the initial profiles are identical.  

In the situation where $\beta_0 > 0$, the initial data is expected to allow the admission of a solution with a soliton component as $t$ increases.  The finite dimensional system shows that if we initially perturb the system either towards the stable or the unstable side of the curve, the system produces immediate oscillations.  Specifically, if the dynamics begin to diverge, the higher order nonlinear corrections in \eqref{quadraticsystem} arrest the solution, resulting in fairly uniform oscillations about the minimal mass soliton.  See Figures \ref{fig:idfdosc1}, \ref{fig:idfdosc2}, \ref{fig:idfdosc3}.  As one can see, for initial values $\omega_0 \approx \omega^*$, we see a good fit for several oscillations of our finite dimensional approximation to the dynamics of the full solution.  As expected, this weakens as $\omega_0$ diverges from $\omega^*$ due to the nature of our approximations in Section \ref{sec:odeder}.  We conjecture that such oscillations about the minimal mass when coupled to the continuous spectrum will lead generically to a damped convergence of the solution towards the minimal mass soliton on a long time scale.

For $\beta_0 < 0$, and other initial parameters sufficiently small, the initial data is below the minimal mass and therefore is expected to disperse as $t$ increases.  In this regime, we see an interesting phenomenon, which is that the motion of our finite dimensional system is a march along the unstable soliton curve towards the value $\omega = 0$.  Clearly, the conservation laws of \eqref{nls} forbid this from happening for the nonlinear solution on long time scales, and we see divergence of the nonlinear solution from our solution as $t$ increases.  However on shorter time scales we see a good fit of the full solution to the predicted finite dimensional dynamics; see Figures \ref{fig:idfdosc4}, \ref{fig:idfdosc5}, \ref{fig:idfdosc6}.  Note that this occurs regardless of whether we initially perturb in the stable or unstable direction.  Our findings confirm the regimes predicted by the dynamical system model in \cite{pelinovsky1996nto}.

It remains to briefly comment on the convergence of our numerical methods.  The stiff ODE solver, {\it ode45}, for the finite dimensional system of ODEs is a standard Runge-Kutta method with known stong convergence results documented in a number of introductory texts on numerical methods.  In addition, the finite element solver for the full nonlinear problem has well-established analytic convergence results, see \cite{ADKM1}.  Hence, the solutions for the corresponding systems are known to be accurate representations of the actual continuous solutions.  Though we have not fully justified in this work the spectral decomposition used to derive \eqref{quadraticsystem}, the fact that the infinite dimensional dynamics are so well approximated by the finite dimensional system constructed from these spectral assumptions is quite good evidence that this approximation is a valid one.  However, as mentioned in Section $2.1$, investigating the validity of spectral assumptions will be an important topic of future research.

\begin{figure}
\includegraphics[width=4.0in]{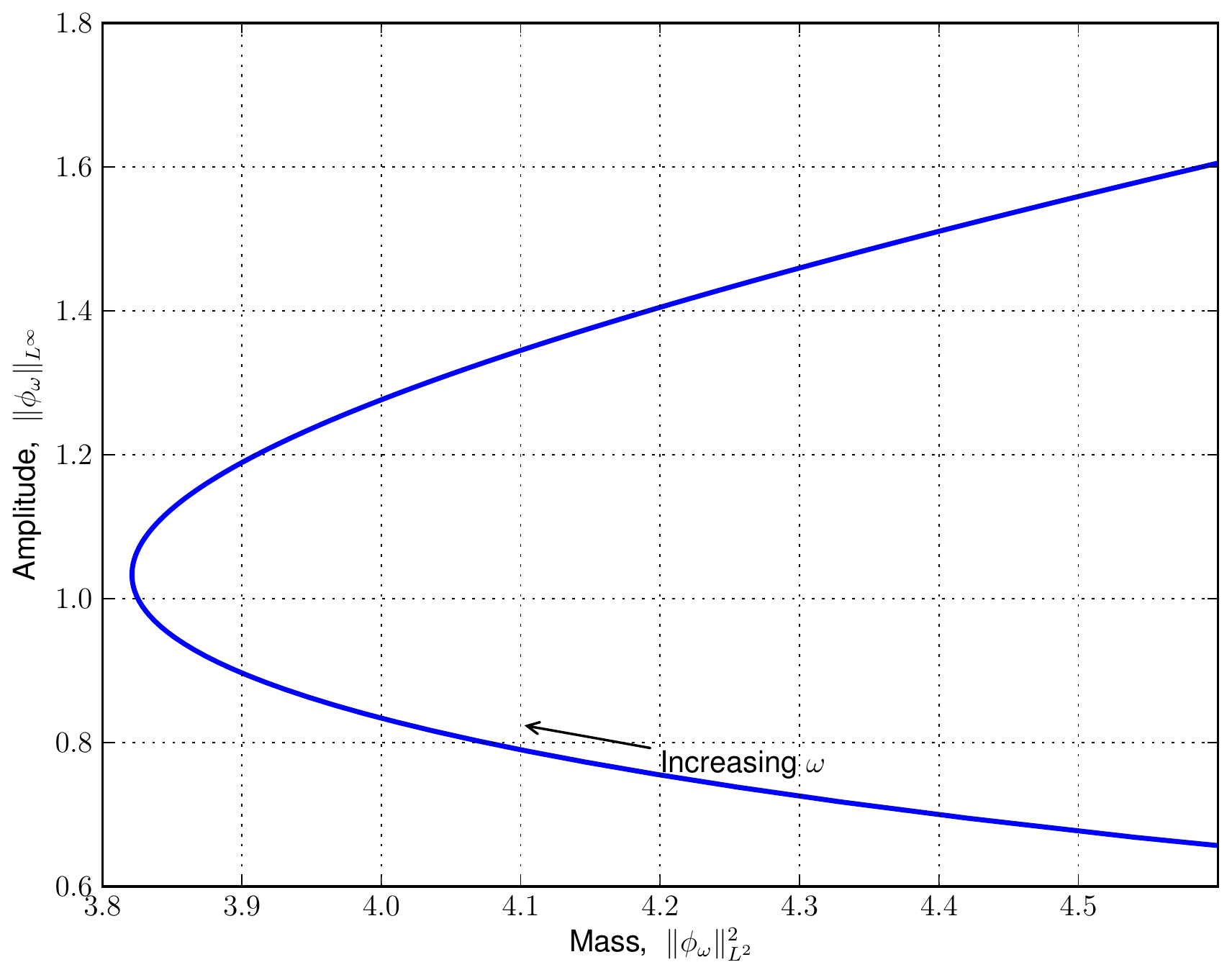}
\caption{A plot of the maximum amplitude with respect to the $L^2$ norm for a saturated nonlinear Schr\"odinger equation.  Computed at $M+1=101$ collocation points for $\omega \in [0.01, 1.5]$.}
\label{fig:nonlinosc}
\end{figure}

\section{Conclusions and Discussion of Future Work}

In this work, we have used a sinc discretization method  to compute the coefficients of the dynamical system \eqref{quadraticsystem}, which is valid near the minimal mass soliton for a saturated nonlinear Schr\"odinger equation.  We find that the dynamical system is an accurate approximation to the full nonlinear solution in a neighborhood of the minimal mass.  Moreover, we see that there are two distinct regimes of the dynamical system.

The first regime represents oscillation along the soliton curve.  The finite dimensional oscillations are valid solutions on long time scales in the conservative PDE, hence we may observe long time closeness of our finite dimensional approximation to the full solution of \eqref{nls}.

The second regime represents a strong forcing in our finite dimensional system towards the point $\omega = 0$ in finite time.  As $\omega \to 0$, the soliton profile becomes small and broad, which is essentially indistinguishable from dispersion.  Hence, it is fitting that we observe these finite dimensional dynamics precisely when the full solution is expected to become completely dispersive.  The strong forcing regime is only valid on a finite time scale, since $\| \phi_\omega \|_{L^2} \to \infty$ as $\omega \to 0$ and \eqref{nls} is conservative.  Our numerical evidence suggests that dispersive dynamics initially move a solution along the unstable portion of the soliton curve, until conservation no longer allows such motion.

We cannot numerically verify our conjecture that soliton preserving perturbations of unstable solitons dynamically select stable solitons.  However, when we begin with perturbations that are expected to continue to have a soliton component, we see oscillations about the minimal mass; this strongly suggests that, through coupling to the continuous spectrum, the oscillations will damp towards a near-minimal-mass stable soliton.  This would be quite satisfying from a physical perspective as the system would be moving towards the configuration of lowest energy in some sense.

In this result, we felt it worthwhile to first understand the underlying finite dimensional dynamics of \eqref{fullsystem1}, even in an asymptotic setting.  However, in the future, we hope to give analytic descriptions of the dynamics of small perturbations of unstable solitons on global or near global time scales by looking at the finite dimensional dynamics coupled to the continuous spectrum dynamics. In the oscillatory regime, this should result in a damped decay of the oscillations to a near minimal mass soliton.  In the strong forcing regime, this should provide a mechanism for mass transfer into the purely dispersive part of the spectrum.  Likely, using current techniques this analysis can only be truly done in a perturbative setting, though we believe that initial conditions near strongly unstable solitons should exhibit similar behavior.  Hopefully more powerful techniques will eventually be developed for the global study of the stable soliton curve as an attractor of the full nonlinear dynamics.

\bibliography{mrs}

\appendix

%\section{Details of Numerical Methods}

\section{Details of Numerical Methods}

\subsection{Sinc Approximation}
\label{app:sinc_details}
Here we briefly review $\sinc$ and its properties.  The texts \cite{lund1992smq, stenger1993nmb} and the articles \cite{stenger1981nmb, bellomo1997nmp, stenger2000ssn} provide an excellent overview.  As noted, sinc collocation and Galerkin schemes have been used to solve a variety of partial differential equations.

Recall the definition of $\sinc$,
%\begin{defn}
%For all $z \in \mathbb{C}$
\begin{equation}
\label{eq:sinc_def}
\sinc(z) \equiv \begin{cases}
\frac{\sin(\pi z)}{\pi z}, &\text{if $z\neq 0$}\\
1, &\text{if $z=0$}
\end{cases} .
\end{equation}
%\end{defn}
and for any $k \in \mathbb{Z}$, $h>0$, let
\begin{equation}
S(k,h)(x) = \sinc\paren{\frac{x-kh}{h}} .
\end{equation}

The sinc function can be used to exactly represent functions in the Paley-Wiener class.  We spectrally represent functions with sinc in a weaker function space.  First, we define a \emph{strip} in the complex plane,
\begin{equation}
\label{eq:stripdomain}
D_d = \set{z\in \mathbb{C}\mid \abs{\Im z}< d} .
\end{equation}
Then we define the function space:
\begin{defn}
\label{def:bp}
$B^p(D_d)$ is the set of analytic functions on $D_d$ satisfying:
\begin{subequations}
\begin{gather}
\norm{f(t+ \mathrm{i} \cdot)}_{L^1(-d,d)} = O(\abs{t}^a),\quad \text{as $t\to \pm \infty$, with $a \in [0,1)$} ,\\
\lim_{y\to d^-} \norm{f(\cdot+ \mathrm{i} y)}_{L^p}+ \lim_{y\to d^-} \norm{f(\cdot- \mathrm{i} y)}_{L^p} <\infty .
\end{gather}
\end{subequations}
\end{defn}

%\begin{equation}
%\label{eq:expansion}
%C_{M}(f,h)(x) = \sum_{k=-M}^{M} f(kh) S(k,h)(x) = \sum_{k=-M}^M f_k S(k,h)(x)
%\end{equation}
Then, we have the following

\begin{thm}  (Theorem 2.16 of \cite{lund1992smq})
\label{thm:sinc_convergence}

Assume $f \in B^p(D_d)$, $p=1$ or $2$, and $f$ satisfies the decay estimate
\begin{equation}
\label{eq:decay}
\abs{f(x)}\leq C \exp{-\alpha \abs{x}} .
\end{equation}
If $h$ is selected such that
\begin{equation}
\label{eq:h_condition}
h = \sqrt{\pi d /(\alpha M)}\leq \min\set{\pi d, \pi/\sqrt{2}} ,
\end{equation}
then
\[
\norm{\partial_x^{n} f -\partial_x^n C_M(f,h)}_{L^\infty}\leq C M^{(n+1)/2}\exp{\paren{(-\sqrt{\pi d\alpha M})}} .
\]
\end{thm}
$d$ identifies a strip in the complex plane, of width $2d$, about the real axis in which $f$ is analytic.  This parameter may not be obvious; others have found $d=\pi/2$ sufficient.

For the NLS equation of order $2\sigma+1$,
\[
d_{\textrm{NLS}} = \frac{\pi}{\sqrt{2 \omega \sigma}} .
\]
Saturated NLS ``interpolates" between second and seventh order NLS.  We thus reason it is fair to take $d = \pi / \sqrt{6 \omega}$.  Though we do not prove that the soliton and the associated elements of the kernel lie in these $B^p(D_d)$ spaces or satisfy the hypotheses of Theorem \ref{thm:sinc_convergence}, we use \eqref{eq:h_condition} to guide our selection of an optimal $h$.  Since the soliton has $\alpha = \sqrt{\omega}$, we reason that it should be acceptable to take
\begin{equation}
\label{eq:h_soliton}
h = \sqrt{\frac{\pi d}{\alpha M}} = \sqrt{\frac{\pi^2 }{6 \omega M}} .
\end{equation}
\eqref{eq:h_soliton} is dependent on both $M$ and $\omega $.  Were we to use \eqref{eq:h_soliton} as is, it would complicate approximating, amongst other things, the derivative with respect to $\omega$ of the soliton.  To avoid this, we use \emph{a priori} estimates on $\omega^\ast$, given in Appendix \ref{sec:quad_root}.  Since we know that $\omega^\ast \sim .18< .25$, it is sufficient to take
\[
d =\pi\sqrt{2/3} .
\]
Likewise, since $\omega^\ast  > .1$, we may take
\[
\alpha = \sqrt{1/10} .
\]
Thus, instead of \eqref{eq:h_soliton}, we use
\begin{equation}
\label{eq:h_prag}
h = \pi \sqrt{\frac{\sqrt{20/3}}{M}} .
\end{equation}
We conjecture that this is a valid grid spacing for all $\omega \in (.1, .25)$; our computations are consistent with this assumption.

\subsection{Numerical Continuation}
\label{sec:continuation}
As discussed in Section \ref{sec:sinc_disc}, the discrete system approximating \eqref{eqn:sol} is
\begin{equation}
\label{eq:discrete_satnls}
\vec{F}(\vec{\phi})= D^{(2)} \vec{\phi} -\omega \vec{\phi} + g(\vec{\phi}) \vec{\phi}=0 .
\end{equation}
The multiplication in $g(\vec{\phi}) \vec{\phi}$ is performed elementwise.  In order to solve this discrete system, we need a good starting point for our nonlinear solver.  We produce this guess by numerical continuation.

Define the function
\[
\hat{g}(x; \tau) = \frac{x^3}{1 + \tau x^2} .
\]
Note that $\hat{g}(x,0)$ is 7th order NLS and $\hat{g}(x,1)=g(x)$, saturated NLS.  We now solve
\begin{equation}
\label{eq:discrete_satnls_mod}
\vec{G}(\vec{\phi};\tau)= D^{(2)} \vec{\phi} -\lambda \vec{\phi} + \hat{g}(\vec{\phi};\tau) \vec{\phi}=0 .
\end{equation}
At $\tau=0$, the analytic NLS soliton serves as the initial guess for computing $\vec{\phi}_{\tau = 0}$.  $\vec{\phi}_{\tau = 0}$ is then the initial guess for solving \eqref{eq:discrete_satnls_mod} at $\tau = \Delta \tau$.  We iterate in $\tau$ until we reach $\tau = 1$.  This is numerical continuation in the artificial parameter $\tau$, \cite{allgower1990ncm}. This process succeeds with relatively few steps of $\Delta \tau$; in fact only $O(10)$ steps are required.

\subsection{Convergence Data}
\label{sec:convergence_data}

Table \ref{table:param_results} offers some examples of the robust and rapid convergence seen in the coefficients of \eqref{quadraticsystem}.  These values are all computed at the minimal mass soliton.  Also see Table \ref{tab:mass_conv}.

\begin{table}
\caption{The convergence of several coefficients for the ODE system, computed at $\omega^\ast$.}
\label{table:param_results}
\begin{tabular}{r| l | l| l|l}
 $M$ & $a_0$ & $c14$ & $p13$ & $n133$ \\
\hline
  20  &  -0.54851448504 &  6.04829942099  & -6.04829942099 & 8.79376331231 \\
  40  &  -0.553577138662 &  6.12521927361  & -6.12521927361 & 8.81625889017 \\
  60  &  -0.553555163933 &  6.12811479039  & -6.12811479039 & 8.81709463059 \\
  80  &  -0.553550441653 &  6.12827391288  & -6.12827391288 & 8.81713847275 \\
 100  &  -0.553549989603 &  6.12828576495  & -6.12828576495 & 8.81714173314 \\
 200  &  -0.553549934797 &  6.12828700415  & -6.12828700415 & 8.81714206225 \\
 300  &  -0.553549934793 &  6.12828700423  & -6.12828700423 & 8.81714206227 \\
 400  &  -0.553549934795 &  6.12828700421  & -6.12828700421 & 8.81714206223 \\
 500  &  -0.553549934794 &  6.12828700423 & -6.12828700423 & 8.81714206227
\end{tabular}
\end{table}

\begin{table}
\caption{Value of the coefficients in \eqref{quadraticsystem} computed with $M=200$.}
\begin{tabular}{r | l}
Coefficient & Value \\
\hline
 $g_{33}$ & -6.61999411752 \\
 $g_{44}$ & -12.4582451458 \\
 $c_{14}$ & 6.12828700415 \\
 $c_{23}$ & 1.46358108488 \\
 $c_{34}$ & 4.0422919871 \\
 $c_{43}$ & 0.131304385722 \\
 $p_{13}$ & -6.12828700415 \\
 $p_{24}$ & -17.9305799071 \\
 $p_{33}$ & 6.61999411752 \\
 $p_{44}$ & 12.4582451458 \\
 $n_{133}$ & 8.81714206225 \\
 $n_{144}$ & 1.84068246508 \\
 $n_{234}$ & 1.45559877602 \\
 $n_{333}$ & -0.792198288158 \\
 $n_{344}$ & 0.013887281387 \\
 $n_{434}$ & -0.0822482271619 \\
 $a_0$ & -0.553549934797
\end{tabular}
\end{table}

%\section{Reference Calculations for the Minimal Mass}
\subsection{Comparisons with Quadrature Methods}
\label{sec:quad_root}
The soliton equation may be integrated once to get
\begin{equation}
\label{eq:first_integral}
\frac{1}{2}(\partial_x\phi)^2 - \frac{1}{2}\lambda \phi ^2 + \frac{1}{4} \bracket{\phi ^4 - \log\paren{1 + \phi ^4}}=0 .
\end{equation}
Equation \eqref{eq:first_integral} yields an implicit algebraic expression for the amplitude, $\phi(0)$,
\begin{equation}
\label{eq:amp}
- \frac{1}{2}\lambda \phi(0) ^2 + \frac{1}{4} \bracket{\phi(0) ^4 - \log\paren{1 + \phi(0) ^4}}=0 .
\end{equation}
Using \eqref{eq:first_integral} and \eqref{eq:amp}, we can express the mass as
\[
\norm{\phi}_{L^2}^2 = \int_{-\infty}^\infty \phi(x)^2 dx = 2 \int_0^\infty \phi(x)^2 dx = 2 \int_0^{\phi(0)}{\rho^2}\set{{\lambda \rho^2 - \frac{1}{2}\bracket{\rho^4 - \log\paren{1+\rho^4}}}}^{-1/2}  d \rho .
\]
Thus, the mass of the soliton with parameter $\omega$ is
\begin{equation}
\label{eq:mass_integral}
\norm{\phi_\omega}_{L^2}^2= 2 \int_0^{\phi(0; \omega)}{\rho^2}\set{{\omega \rho^2 - \frac{1}{2}\bracket{\rho^4 - \log\paren{1+\rho^4}}}}^{-1/2}  d \rho .
\end{equation}
Equations \eqref{eq:amp} and \eqref{eq:mass_integral} can be used to approximate $\omega^\ast$ by numerically minimizing \eqref{eq:mass_integral}.  To compute the amplitude of the soliton, we solve \eqref{eq:amp} using Brent's method with a tolerance of 1.0e-14. We use the singular integral integrator QAGS from QUADPACK, which for this problem is, unfortunately, limited to a relative error of 5.0e-12 and an absolute error of 1.0e-15.  Trying different routines from the optimization module of SciPy, \cite{jones2001},  we summarize our results in Table \ref{table:quadmin_results}.  There is a spread of O(1e-12) amongst the computed minimal masses and a spread of O(1e-7) amongst the $\omega^\ast$.  These differences are consistent with the prescribed relative error of the quadrature, suggesting the precision of this approach to computing the minimal mass and associated $\omega$ is limited by the quadrature algorithm.  We summarize these computations in Table \ref{table:quadmin_results}, which also contains data from our $\sinc$ computations.    The $\sinc$ method is consistent with these quadrature methods.

%This requires us to:
%\begin{enumerate}
%\item Compute the amplitude of the soliton $R(x; \lambda)$ by numerically finding the roots of \eqref{eq:amp}
%\item Numerically integrate \eqref{eq:mass_integral}.
%\end{enumerate}

\begin{table}
\caption{The soliton parameter and mass of the minimal mass soliton computed by quadrature.  Also included is some of the data for the $\sinc$ method appearing in Table \ref{tab:mass_conv}.  }
\label{table:quadmin_results}
\begin{tabular}{r|l|l}
Algorithm & $\omega^\ast$ & $\int \abs{\phi}^2 dx$\\
\hline
fminbound & 0.177588368745261 &3.821149022780204 \\
Brent & 0.177587963826864 & 3.821149022778472\\
golden & 0.177587925853761 &3.821149022776717\\
$\sinc$ with $M= 100$ & 0.177588063805561& 3.821149022493814 \\
$\sinc$ with $M = 200$ & 0.177588065432740 & 3.821149022780618 \\
$\sinc$ with $M = 400$ & 0.177588065433095 & 3.821149022780896
\end{tabular}
\end{table}

\end{document}